\documentclass[reprint,prl,twocolumn,superscriptaddress,showpacs,showkeys,floatfix,preprintnumbers]{revtex4-2}

\usepackage{graphicx}
\usepackage{dcolumn}
\usepackage{bm}
\usepackage{hyperref}

\usepackage{braket}
\usepackage{xargs}
\usepackage{mathtools}
\usepackage[english]{babel}
\usepackage[utf8]{inputenc}
\usepackage{diagbox}
\usepackage{amsmath,amssymb,braket,slashed,mathrsfs}
\usepackage{pifont}
\usepackage{MnSymbol}
\usepackage{graphicx}
\graphicspath{{./fig/}}
\usepackage{tikz}
\usepackage{color,hyperref}
\usepackage{multirow,array}
\usepackage{textcomp}

    \newcommand{\ba}{\begin{eqnarray}}
    \newcommand{\ea}{\end{eqnarray}}
    \newcommand{\be}{\begin{equation}}
    \newcommand{\ee}{\end{equation}}
    \newcommand{\nn}{\nonumber}

    \newcommand{\innovation}{Collaborative Innovation Center of Quantum Matter, Beijing 100871, China} 
    \newcommand{\chep}{Center for High Energy Physics, Peking University, Beijing 100871, China}    
    \newcommand{\pkuphy}{School of Physics, Peking University, Beijing 100871,
	China}
    \newcommand{\Uconn}{Department of Physics, University of Connecticut, Storrs, CT 06269, USA}
    \newcommand{\Regensburg}{Fakult\"at f\"ur Physik, Universit\"at Regensburg, Universit\"atsstra{\ss}e 31, 93040 Regensburg, Germany}
    \newcommand{\Milano}{Dipartimento di Fisica, Universit\`adi Milano-Bicocca, Piazza della Scienza 3, I-20126 Milano, Italy}
    \newcommand{\INFN}{INFN, Sezione di Milano-Bicocca, Piazza della Scienza 3, I-20126 Milano, Italy}

\begin{document}

\title{Lattice QCD calculation of the $\pi^0$-pole contribution to the hadronic light-by-light
scattering in the anomalous magnetic moment of the muon}

	\author{Tian~Lin}\affiliation{\pkuphy}
        \author{Mattia~Bruno}\affiliation{\Milano}\affiliation{\INFN}
        \author{Xu~Feng}\affiliation{\pkuphy}\affiliation{\innovation}\affiliation{\chep}
        \author{Lu-Chang Jin}\affiliation{\Uconn}
        \author{Christoph~Lehner}\affiliation{\Regensburg}
        \author{Chuan~Liu}\affiliation{\pkuphy}\affiliation{\innovation}\affiliation{\chep}
	\author{Qi-Yuan~Luo}\affiliation{\pkuphy}

\date{\today}

\begin{abstract}

We develop a method to compute the pion transition form factor directly at arbitrary space-like photon momenta and use it to determine 
the $\pi^0$-pole contribution to the hadronic light-by-light scattering in the anomalous magnetic moment of the muon.
The calculation is performed using eight gauge ensembles generated with 2+1 flavor domain wall fermions, incorporating multiple pion masses, lattice spacings, and volumes.
By introducing a pion structure function and performing a Gegenbauer expansion, we demonstrate that about 98\% of the $\pi^0$-pole contribution can be extracted in a model-independent manner, thereby 
ensuring that systematic effects are well controlled. After applying finite-volume corrections, as well as performing chiral and continuum extrapolations, we obtain the final result for the $\pi^0$-pole contribution to 
the hadronic light-by-light scattering in the muon’s anomalous magnetic moment, $a_{\mu}^{\pi^0\mathrm{-pole}}=61.2(1.7)\times 10^{-11}$, and the $\pi^0$ decay width, $\Gamma_{\pi^0\to \gamma\gamma}=7.60(27)$ eV.

\end{abstract}

\maketitle


\section{Introduction}

The anomalous magnetic moment of the muon, $a_\mu$, is one of the most precisely measured quantities in physics. 
The Muon $g–2$ collaboration at Fermilab has recently released their final result, $a_\mu^{\mathrm{exp}} = 116592071(15)\times 10^{-11}$~\cite{Muong-2:2025xyk}, which is consistent with their earlier measurements~\cite{Muong-2:2023cdq,Muong-2:2021ojo}
and the previous Brookhaven result~\cite{Muong-2:2006rrc}.
Notably, this experimental value agrees well with the Standard Model prediction reported earlier in the 2025 white paper by the Muon $g–2$ Theory Initiative~\cite{Aliberti:2025beg},
which gives $a_\mu^{\mathrm{SM}} = 116592033(62)\times 10^{-11}$. This agreement marks another significant success of the Standard Model.
It is important to highlight that the 2025 white paper prediction differs substantially from the one presented in the 2020 white paper~\cite{Aoyama:2020ynm}, which exhibited a 5.7 $\sigma$ tension with the latest experimental result. 
Thanks to considerable progress in lattice QCD calculations and recent developments in data driven analysis,
the 2025 white paper~\cite{Aliberti:2025beg} adopts a lattice-only determination of the hadronic vacuum polarization (HVP) contribution to $a_\mu$, 
drawing on results from a broad set of collaborations~\cite{Bazavov:2024eou, MILC:2024ryz, ExtendedTwistedMass:2024nyi, Djukanovic:2024cmq, RBC:2024fic, Spiegel:2024dec, Boccaletti:2024guq, Kuberski:2024bcj, RBC:2023pvn, FermilabLatticeHPQCD:2023jof, FermilabLattice:2022izv, ExtendedTwistedMass:2022jpw, Ce:2022kxy, Aubin:2022hgm, Wang:2022lkq, Lehner:2020crt, Borsanyi:2020mff, Aubin:2019usy, RBC:2018dos}. 
Compared to the lattice-QCD average used in 2020, the uncertainty of the new average has been reduced by a factor of about 3.
Given these developments, it is prudent to also scrutinize the other major source of theoretical uncertainty in the Standard Model prediction of muon $g-2$, namely the hadronic light-by-light (HLbL) scattering contribution.

\begin{figure}[htbp!]
    \centering
    \includegraphics[width=0.48\textwidth]{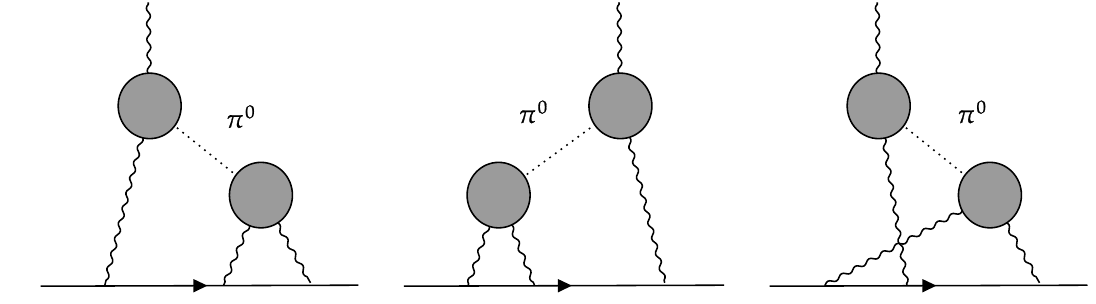}
    \caption{Pion-pole contribution to the HLbL.}
    \label{fig:pion-pole}
\end{figure}

For HLbL, two complementary approaches yield predictions that are in small tension with each other, at the level of about 1.5 $\sigma$~\cite{Aliberti:2025beg}.
One involves direct lattice QCD computations of the HLbL contribution using vector-current four-point correlation functions~\cite{Blum:2014oka,Blum:2015gfa,Blum:2016lnc,Blum:2017cer,Chao:2021tvp,Chao:2022xzg,Asmussen:2022oql,Blum:2023vlm,Fodor:2024jyn}.
The other is a data-driven analysis based on a dispersive framework~\cite{Colangelo:2014dfa,Colangelo:2014pva,Pauk:2014rfa,Colangelo:2015ama,Hoferichter:2024vbu,Hoferichter:2024bae,Holz:2024diw,Holz:2024lom,Bijnens:2024jgh,Bijnens:2022itw,Bijnens:2021jqo,Bijnens:2020xnl,Bijnens:2019ghy}, where the leading hadronic contribution arises from pseudoscalar poles, particularly the $\pi^0$ pole (see Fig.~\ref{fig:pion-pole}). 
Several studies have calculated the pseudoscalar-pole contribution to HLbL using lattice QCD~\cite{Gerardin:2016cqj,Gerardin:2019vio,Gerardin:2023naa,ExtendedTwistedMass:2023hin,Koponen:2023zle,ExtendedTwistedMass:2022ofm}.
So far, both approaches yield comparable uncertainties. Since the $\pi^0$-pole contribution amounts for $2/3$ of the total HLbL, a careful examination of systematic effects and further reduction of the total uncertainties will be crucial for refining the theoretical prediction of HLbL.

The $\pi^0$-pole contribution to HLbL, $a_{\mu}^{\pi^0\mathrm{-pole}}$, is defined by the dispersive formula~\cite{Jegerlehner:2009ry}
\begin{eqnarray}
\label{eq:amu-pion-pole}
    &&a_{\mu}^{\pi^0\mathrm{-pole}}=\left(\frac{\alpha}{\pi}\right)^3\int_{0}^{\infty}dQ_1\int_{0}^{\infty}dQ_2\int_{-1}^{1}d\tau \nonumber\\
    &&\hspace{0.3cm} w_1(Q_1,Q_2,\tau) \mathcal{F}_{\pi^0\gamma^*\gamma}(-Q_1^2,0)\mathcal{F}_{\pi^0\gamma^*\gamma^*}(-Q_2^2,-Q_3^2)+ \nonumber\\
     &&\hspace{0.3cm} w_2(Q_1,Q_2,\tau) \mathcal{F}_{\pi^0\gamma^*\gamma}(-Q_3^2,0)\mathcal{F}_{\pi^0\gamma^*\gamma^*}(-Q_1^2,-Q_2^2),
\end{eqnarray}
where $w_1,w_2$ are known weight functions, 
and $Q_3$ is defined in terms of $\tau$ with $\tau\equiv\frac{Q_3^2 - Q_1^2-Q_2^2}{2Q_1Q_2}$. The key component here is
the $\pi^0$ transition form factor (TFF), $\mathcal{F}_{\pi^0\gamma^*\gamma^*}(-Q^2,-{Q'}^2)$, which is essential for calculating $a_{\mu}^{\pi^0\mathrm{-pole}}$. 
Previous studies~\cite{Feng:2012ck,Gerardin:2016cqj,Gerardin:2019vio,Gerardin:2023naa,ExtendedTwistedMass:2022ofm,ExtendedTwistedMass:2023hin,Koponen:2023zle} adopt the time-momentum representation
as proposed by Ref.~\cite{Ji:2001wha}, where momenta for the pion and photons are limited to discrete values, necessitating parametrization like the $z$-expansion for momentum extrapolation.
However, the errors from the truncation of the $z$-expansion might be hard to quantify.
In practical lattice calculations, 
typically 4-6 parameters are used to fit the entire TFF. While it can be demonstrated that the lattice data are consistent with the fit, ensuring that the results are constrained by the lattice data 
rather than by the fitting form is challenging, especially in momentum regions where direct lattice calculations are noisier than the fit. 
To mitigate this, we adopt a different approach based on a theoretical constraint derived in this work (Eq.~(\ref{eq:structure_func})), 
allowing us to calculate the majority of the TFF in a completely model-independent manner. 
The remaining small model dependent part is isolated, so we can study the dependence explicitly.

Our calculation of TFF with arbitrary space-like photon momenta builds upon a technique previously employed in the studies 
of $\pi^0\to e^+e^-$~\cite{Christ:2022rho} and $\eta_c\to\gamma\gamma$~\cite{Meng:2021ecs}.
Applying this method to calculate $a_{\mu}^{\pi^0\mathrm{-pole}}$, however, introduced new challenges, necessitating further methodological developments and key refinements. Notably, we observed that the quark-disconnected diagrams contribute with the same sign as the quark-connected ones, contrary to previous results~\cite{Feng:2012ck, Gerardin:2016cqj, Gerardin:2019vio, Gerardin:2023naa, ExtendedTwistedMass:2023hin, Koponen:2023zle, Christ:2022rho}. This finding partly clarifies the earlier 1-2\,$\sigma$ tension between lattice calculations and experimental measurements of the $\pi^0$ decay width.

\section{Methodology}

We begin by calculating the $\pi^0\to\gamma^*\gamma^*$ TFF in infinite
    volume. The relevant
    hadronic matrix element in Euclidean spacetime is given as
        \be
\mathcal{H}_{\mu\nu}(x) \equiv \left\langle 0\left|T\left\{J_{\mu}\left(\frac{x}{2}\right)J_{\nu}\left(-\frac{x}{2}\right)\right\}\right|\pi(P)\right\rangle,
        \ee
        where $J_\mu(x)$ represents the electromagnetic current, and $P=(iE_\pi,\vec{p})$ denotes the
        4-momentum of the on-shell pion. The TFF can be extracted via the Fourier transform
\be
\label{eq:F_munu}
\mathcal{F}_{\mu\nu}(Q,Q')= \int d^4x\, e^{-i\left(Q-\frac{P}{2}\right)\cdot x}\mathcal{H}_{\mu\nu}(x),
\ee
with $Q'=P-Q$ and $Q=(iE,\vec{q})$ taking arbitrary real values for $E$ and $\vec{q}$.
The hadronic function $\mathcal{H}_{\mu\nu}(x)$ in coordinate space and
$\mathcal{F}_{\mu\nu}(Q,Q')$ in momentum space
 share the same Lorentz structure and can be decomposed as
\ba
\mathcal{H}_{\mu\nu}(x) &=& -\varepsilon_{\mu\nu\alpha\beta}x_{\alpha}P_{\beta} H(x^2,P\cdot x),
\nn\\
\mathcal{F}_{\mu\nu}(Q,Q')&=&i\varepsilon_{\mu\nu\alpha\beta}Q_{\alpha}P_{\beta}\mathcal{F}_{\pi^0\gamma^*\gamma^*}(-Q^2,-{Q'}^2).
\ea
The relation between the scalar functions $H(x^2,P\cdot x)$ and $\mathcal{F}_{\pi^0\gamma^*\gamma^*}(-Q^2,-{Q'}^2)$ is given in
\ba
\label{eq:relation}
        &&\mathcal{F}_{\pi^0\gamma^*\gamma^*}(-Q^2,-Q'^2) = \int d^4x \,\omega(K,P,x) H(x^2,P\cdot x),
        \nn\\
        &&\omega(K,P,x) \equiv i\,e^{-iK\cdot x} \frac{(K\cdot x) P^2 - (K\cdot P)(P\cdot x)}{K^2P^2-(K\cdot P)^2},
\ea 
where $K=Q-P/2$.

In the pion rest frame, where $P = (im_{\pi},0)$, $H(x^2,P\cdot x)$ remains invariant under spatial rotations. 
By performing the spatial $SO(3)$ average of $\omega(K,P,x)$, we get
\be
\label{eq:weight_function_SO3}
\langle \omega(K,P,x)\rangle_{SO(3)}=\,e^{(E-\frac{1}{2}m_{\pi})t}\frac{|\vec{x}|}{|\vec{q}|}j_1(|\vec{q}||\vec{x}|),
\ee
where $j_{1}(x)$ represents a spherical Bessel function. 
While the factor $e^{(E - \frac{1}{2}m_{\pi})t}$ grows exponentially, the rapid oscillation of $j_1(|\vec{q}||\vec{x}|)$ causes exponential suppression in the spatial integral,
ensuring convergence of the spacetime integral in Eq.~(\ref{eq:relation}) for accurate TFF.
However, in practical lattice calculations, numerical evaluation of $H(x^2, P \cdot x)$ introduces statistical noise. 
Although the signal cancels out in the integral, the noise does not, leading to significant uncertainty in the lattice data, particularly when $E$ is large~\cite{SM}.

The challenge lies in the fact that both the hadronic function $H(x^2, P\cdot x)$ and the weight function are only $SO(3)$-symmetric. 
Achieving $SO(4)$-symmetry in the hadronic function would greatly simplify the weight function.
To address this issue, we introduce a pion structure function $\phi_{\pi}(x^2, u)$ and express $H(x^2, P \cdot x)$ as
\be
\label{eq:structure_func}
H(x^2, P\cdot x) = \int_{0}^{1}du\, e^{i\left(u-\frac{1}{2}\right)P\cdot x}\phi_{\pi}(x^2,u)H(x^2,0),
\ee
where $\phi_{\pi}(x^2,u)$ satisfies the normalization condition $\int_{0}^{1} du\, \phi_{\pi}(x^2, u) = 1$, $\forall\, x^2$. 
Expressions like Eq.~(\ref{eq:structure_func}) typically arise in perturbative analyses.
At small spacelike separations, i.e., $x^2 \ll \Lambda_{\mathrm{QCD}}^{-2}$, $\phi_{\pi}(x^2, u)$ corresponds to the pion distribution amplitude, up to $\mathcal{O}(\alpha_s)$ corrections and higher-twist effects~\cite{Braun:2007wv,Bali:2017gfr,Bali:2018spj}.
In the Supplemental Material~\cite{SM}, we demonstrate  that  Eq.~(\ref{eq:structure_func}) also holds in the non-perturbative regime, implying that $\phi_{\pi}(x^2,u)$ 
represents more than just a pion distribution amplitude.
The symmetry $\phi_{\pi}(x^2, u) = \phi_{\pi}(x^2, 1 - u)$ follows from the property $\mathcal{H}_{\mu\nu}(x)=\mathcal{H}_{\nu\mu}(-x)$.

Substituting Eq.~(\ref{eq:structure_func}) into Eq.~(\ref{eq:relation}), we obtain
\ba
\label{eq:relation_SO4}
\mathcal{F}_{\pi^0\gamma^*\gamma^*}(-Q^2,-Q'^2)& =& \int_0^1 du\int d^4x \,\omega(\bar{K},P,x)
\nn\\
 &&\hspace{0.5cm}\times\phi_{\pi}(x^2,u)H(x^2,0),
\ea
with $\bar{K}=Q-uP$. Since $\phi_{\pi}(x^2,u)H(x^2, 0)$ is symmetric under $SO(4)$ spacetime rotations, we can perform an $SO(4)$ average for the weight function $\omega(\bar{K}, P, x)$
\be
\label{eq:weight_function_SO4}
\langle \omega(\bar{K},P,x)\rangle_{SO(4)}=2\frac{J_2(|\bar{K}||x|)}{\bar{K}^2},
\ee
with $J_2(x)$ the Bessel function. Notably, for the $SO(4)$ average, the pion need not be at rest, and the formulas are valid for any momentum $P$. Since $J_2(|\bar{K}||x|)$ does not 
exhibit exponential growth at large $|\bar{K}||x|$, there is no signal-to-noise problem in computing the TFF.
To fully utilize the lattice data, we replace $H(x^2, 0)$ in Eq.~(\ref{eq:relation_SO4}) with $H(x^2, P \cdot x)$ from Eq.~(\ref{eq:structure_func}), resulting in the master formula
\ba
\label{eq:master_formula}
&&\mathcal{F}_{\pi^0\gamma^*\gamma^*}(-Q^2,-Q'^2)
\nn\\
&=&\frac{2}{\eta(P)}\int d^4x\int_0^1 du \,\frac{J_2(|\bar{K}||x|)}{\bar{K}^2}\phi_{\pi}(x^2,u)\frac{x_\perp}{|x|}H(x^2,0)
\nn\\
&=&\frac{2}{\eta(P)}\int d^4x\frac{\int_0^1 du \,\frac{J_2(|\bar{K}||x|)}{\bar{K}^2}\phi_{\pi}(x^2,u)\frac{x_\perp}{|x|}H(x^2,P\cdot x)}{\int_0^1 du\, e^{i\left(u-\frac{1}{2}\right)P\cdot x}\phi_{\pi}(x^2,u)},
\nn\\
\ea
where
$\frac{x_\perp}{|x|}\equiv \frac{|x-\frac{P\cdot x}{P^2}P|}{|x|}=\sqrt{1-\frac{(P\cdot x)^2}{P^2x^2}}$, and
$\eta(P)=\frac{1}{2\pi^2}\int d\Omega_x\,\frac{x_\perp}{|x|}$.
$d\Omega_x$ is the solid angle associated with the four-dimensional vector $x$.
In the pion rest frame, we have $\frac{x_\perp}{|x|}=\frac{|\vec{x}|}{|x|}$ and $\eta(P)=\frac{8}{3\pi}$.
For a given $x^2$, the hadronic input $H(x^2, P \cdot x)$ is noisier at small $x_\perp$, so multiplying by $\frac{x_\perp}{|x|}$ emphasizes more accurate data in the integral. 
Compared to Eq.~(\ref{eq:relation_SO4}), using the full data set for $H(x^2, P \cdot x)$ 
reduces the error by 30\%, and including the $\frac{x_\perp}{|x|}$ factor cuts it by an additional 10\%.

The determination of TFF using Eq.~(\ref{eq:master_formula}) depends on the input of $\phi_{\pi}(x^2, u)$, which is extracted from $H(x^2, P \cdot x)$ via Eq.~(\ref{eq:structure_func}), 
presenting an inverse problem.
Fortunately, our primary goal is to calculate $a_{\mu}^{\pi^0\mathrm{-pole}}$, a weighted integral of the TFF, rather than determining them at specific momenta.
This quantity is mainly influenced by low-energy contributions linked to the muon or pion mass scale, making it relatively insensitive to the specifics of $\phi_{\pi}(x^2, u)$. To investigate this further, 
we perform a Gegenbauer expansion for $\phi_{\pi}(x^2,u)$~\cite{Radyushkin:1977gp,Lepage:1979zb,Lepage:1980fj}
\be
\label{eq:Gegenbauer}
\phi_{\pi}(x^2,u) = 6u(1-u) \sum_{n=0}^{\infty} \varphi_{2n}(r) C_{2n}^{(\frac{3}{2})}(2u-1),
\ee
where $r\equiv |x|$ and $C_{2n}^{(\frac{3}{2})}(u)$ are the Gegenbauer polynomials. 
Under this expansion, $a_{\mu}^{\pi^0\mathrm{-pole}}$ can be expressed as
\ba
\label{eq:a_2n_2m}
a_{\mu}^{\pi^0\mathrm{-pole}}&=&\int dr_1\int dr_2\sum_{n\ge m}c_{2n,2m}(r_1,r_2)\varphi_{2n}(r_1)\varphi_{2m}(r_2)
\nn\\
&=&a_{0,0}^{\pi^0\mathrm{-pole}}+a_{2,0}^{\pi^0\mathrm{-pole}}+a_{2,2}^{\pi^0\mathrm{-pole}}+a_{4,0}^{\pi^0\mathrm{-pole}}\cdots.
\ea
Here the coefficients $c_{2n,2m}(r_1,r_2)$ can be computed using the lattice hadronic function as input. 
Given the normalization condition of $\phi_{\pi}(x^2,u)$, we have 
$\varphi_{0}(r)=1$, $\forall\,r$. The structural dependence is captured by $\varphi_{2n}(r) $ for $n>0$. Therefore, the first term, $a_{0,0}^{\pi^0\mathrm{-pole}}$, 
is independent of the specific details of $\phi_\pi(x^2,u)$. In the Supplemental Material~\cite{SM} we demonstrate that $a_{0,0}^{\pi^0\mathrm{-pole}}$ provides the dominant contribution, 
and that the terms $a_{2n,2m}^{\pi^0\mathrm{-pole}}$ with $n$ or $m>0$ 
become rapidly suppressed as $n$ and $m$ increase, assuming $|\varphi_{2n}(r)|$ do not significantly exceed 1.

\section{Numerical analysis}
We utilize 2+1-flavor domain wall fermion gauge ensembles generated by the RBC-UKQCD Collaboration, with parameters listed in Table~\ref{tab:ensemble}~\cite{RBC:2014ntl}.
The primary analysis focuses on the two finer ensembles at near-physical pion masses, 48I and 64I, which use the Iwasaki action. Additionally, we include 
six other ensembles—24D, 32D, 32Df and 24DH (Iwasaki + DSDR action), and 24IH and 32IcH (Iwasaki action)—to control systematic errors.

\begin{table}[htbp!]
    \centering
    \begin{tabular}{c c c c c c c}
        \hline
        Ensembles & $a^{-1}$[GeV] & $L$[fm] & $m_{\pi}$[MeV] &  $N_{\mathrm{conf}}\times N_{\mathrm{meas}}$\\
        \hline\hline
        24D & 1.023(2) & 4.6 & 142.6(3) & $253\times1024$ \\
        32D  & 1.023(2) & 6.2 & 142.5(4) & $63\times2048$ \\
        32Df & 1.378(5) & 4.6 & 142.9(7) & $69\times 1024$ \\
        48I &  1.731(3) & 5.5 & 139.7(3) & $112\times 2048$ \\
        64I & 2.355(5) & 5.4  & 139.0(3) & $119\times2048$ \\
        24DH & 1.023(2) & 4.6 & 343.5(8) &  $37\times 1024$ \\
        24IH & 1.785(5)  & 2.7 & 341.1(1.3) &  $77\times 512$ \\
        32IcH & 1.785(5)  & 3.5 & 340.1(1.2) &  $38\times 512$ \\
        \hline
    \end{tabular}
    \caption{Parameters of the gauge ensembles used in this study. For each ensemble, we provide the inverse of lattice spacing $a^{-1}$, the spatial extent $L$, the pion mass $m_\pi$, and the number of configurations $N_{\mathrm{conf}}$ multiplied by the number of measurements per configuration $N_{\mathrm{meas}}$.}
    \label{tab:ensemble}
\end{table}

We compute the three-point correlation function $\langle J_{\mu}(x) J_{\nu}(y) \bar{O}_{\pi}(t_i)\rangle$
with $t_i = \min(t_x,t_y)-\Delta t$. The time separation $\Delta t$ between the wall-source operator $\bar{O}_{\pi}$ and the nearest electromagnetic current is chosen to be sufficiently large to suppress excited-state contamination. 
More details on the choice of $\Delta t$ and the study of excited-state contamination are 
provided in the Supplemental Material~\cite{SM}.
For the two vector currents, one is treated as the source and the other as the sink, with the source placed at 512, 1024 or 2048 random spacetime locations, denoted by $N_{\mathrm{meas}}$ in Table~\ref{tab:ensemble}. 
The two-point correlation function $\langle O_{\pi}(t)\bar{O}_{\pi}(0)\rangle$ is also computed to provide the normalization factor and pion mass necessary for extracting the matrix element $\mathcal{H}_{\mu\nu}(x)$.  
The Dirac matrix inversions are accelerated using the locally-coherent Lanczos algorithm~\cite{Clark:2017wom}. Both connected and disconnected diagrams are evaluated.

\begin{figure}[htbp!]
    \centering
    \includegraphics[width=0.44\textwidth]{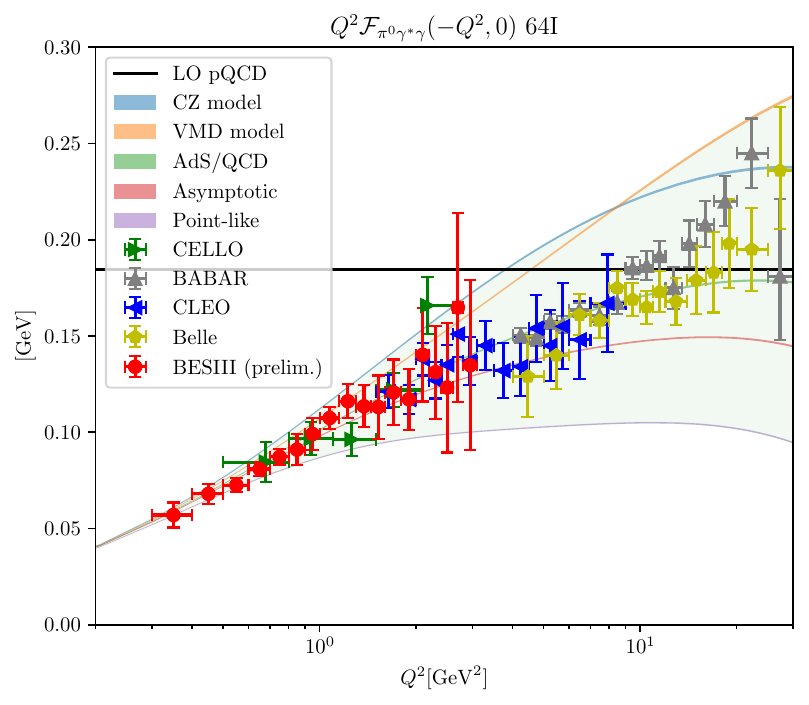}
    \caption{Comparison between experimental measurements of TFF with a single virtual photon and lattice QCD calculations using different parameterizations of $\phi_\pi(x^2,u)$ as inputs.}
     \label{fig:Fq0_ex}
\end{figure}

To study the dependence on the coefficients $\varphi_{2n}$, we introduce five commonly used parameterizations of  $\phi_\pi(x^2,u)$ from the literature~\cite{Lepage:1979zb,Lepage:1980fj,Chernyak:1981zz,RuizArriola:2002bp,Polchinski:2002jw,Brodsky:2003px,Bali:2017gfr,Bali:2018spj,Braun:2007wv}
\ba
\label{eq:model}
        &&\phi_\pi^{\mathrm{VMD}}(x^2,u) = 1,
        \nn\\
        &&\phi_\pi^{\mathrm{AdS/QCD}}(x^2,u) = \frac{8}{\pi}\sqrt{u(1-u)},
        \nn\\
        &&\phi_\pi^{\mathrm{asym}}(x^2,u) = 6u(1-u),
        \nn\\
        &&\phi_\pi^{\mathrm{point-like}}(x^2,u)  = \delta(u-\frac{1}{2}),
        \nn\\
        &&\phi_\pi^{\mathrm{CZ}}(x^2,u)  = 30u(1-u)(1-2u)^2.
\ea
Here, the parameter $x^2$ serves as a scale indicator. As the structure function $\phi_\pi(x^2,u)$ introduced in Eq.~(\ref{eq:structure_func}) 
is valid in both perturbative and non-perturbative regimes, 
we adopt a range of representative forms for $\phi_\pi(x^2,u)$ to estimate its impact. For instance, $\phi_\pi^{\mathrm{VMD}}(x^2,u)$ models the behavior of $\phi_\pi(x^2,u)$ at long distances, 
while $\phi_\pi^{\mathrm{asym}}(x^2,u)$ reflects its asymptotic behavior at short distances.
Using the parameterizations of $\phi_\pi(x^2,u)$ and lattice QCD inputs for $H(x^2,P\cdot x)$, we compute the TFF via Eq.~(\ref{eq:master_formula}). 
Results for a single virtual photon are shown in Fig.~\ref{fig:Fq0_ex}.
The curves from the five parameterizations form a broad region that encompasses all experimental data measured by BABAR~\cite{BaBar:2009rrj}, CELLO~\cite{CELLO:1990klc}, CLEO~\cite{CLEO:1997fho}, Belle~\cite{Belle:2012wwz} and BESIII~\cite{Redmer:2018uew}.
At large momentum transfer,
the TFF can be derived by expanding the product of the two currents on the light-cone, yielding the leading-order prediction in perturbative QCD (LO pQCD) and at leading twist~\cite{Lepage:1979zb,Lepage:1980fj}
\be
\mathcal{F}_{\pi^0\gamma^*\gamma^*}(-Q^2,-{Q'}^2)=\frac{2}{3}F_\pi\int_0^1du\,\frac{\phi_\pi^{\mathrm{asym}}(x^2,u)}{uQ^2+(1-u){Q'}^2},
\ee
where $F_\pi$ is the pion decay constant. As $Q^2\to\infty$, we have $Q^2\mathcal{F}_{\pi^0\gamma^*\gamma}(-Q^2,0)=2 F_\pi$. This LO pQCD prediction, known as Brodsky-Lepage limit, is shown as the black line in Fig.~\ref{fig:Fq0_ex}. 
This line reflects the large-momentum behavior of the TFF, making it challenging to match using parameterizations based on low-momentum expansions, such as the $z$-expansion, without constraints from the high-momentum region.
Nevertheless, it lies safely within the region generated by the five parameterizations mentioned above.
Overall, these results suggest that combining the five parameterizations offers a conservative estimate of the systematic uncertainty introduced by the structure function.

\begin{figure}[htbp!]
    \centering
    \includegraphics[width=0.44\textwidth]{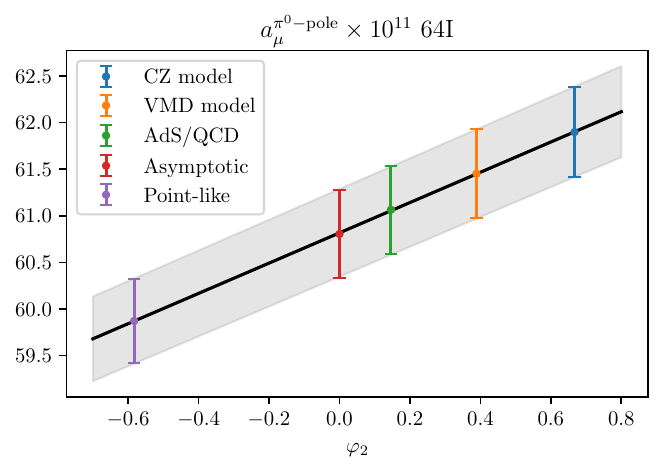}
    \caption{$a_{\mu}^{\pi^0\mathrm{-pole}}$ as a function of $\varphi_{2}(r)$. The results align closely with a linear form, indicating that higher-order effects are negligible.}
    \label{fig:varph2_dependence}        
     \end{figure}

By performing the Gegenbauer expansion of $\phi_\pi(x^2,u)$, we obtain $\varphi_{2}(r)=0.667$, $0.389$, $0.146$, $0$ and $-0.583$ for CZ, VMD, AdS/QCD, OPE asymptotic and point-like parameterizations, respectively. The absolute values of these terms are all smaller than $\varphi_{0}=1$.  As demonstrated in Supplemental Material~\cite{SM}, when these terms contribute to $a_{\mu}^{\pi^0\mathrm{-pole}}$, 
their effects are suppressed by more than an order of magnitude. Higher-order terms, like $\varphi_{4}$, are further suppressed by an additional order of magnitude and can therefore be neglected.
In Fig.~\ref{fig:varph2_dependence}, we illustrate using the 64I ensemble that $a_{\mu}^{\pi^0\mathrm{-pole}}$ depends almost linearly on $\varphi_{2}(r)$, confirming that the contributions from higher-order terms are minimal.

In Eq.~(\ref{eq:model}), the parameterizations of $\phi_\pi(x^2,u)$ are all independent of the scale $x^2$. Consequently, $\varphi_{2}(r)$ is also independent of $r=|x|$. 
When considering the $r$-dependence of $\varphi_{2}(r)$, one can express $a_{2,0}^{\pi^0\mathrm{-pole}}$ in Eq.~(\ref{eq:a_2n_2m}) as
$a_{2,0}^{\pi^0\mathrm{-pole}}=\int dr\, d_{2,0}(r)\varphi_2(r)$,
where $d_{2,0}(r)=\int dr'\,c_{2,0}(r,r')$.
We have calculated $d_{2,0}(r)$ in the Supplemental Material~\cite{SM} and find it to be consistently positive.
As long as $\varphi_2(r)$ falls within the range spanned by the five parameterizations for any $r > 0$, these $x^2$-independent parameterizations for $\phi_\pi(x^2,u)$ will provide a safe bound for the final result.

\begin{table}[htbp!]
    \centering
    \begin{tabular}{c c c c c c c}
        \hline
        Ensembles &  $a_{\mu}^{\pi^0\mathrm{-pole}}\times 10^{11}$\\
        \hline
        24D & $65.65(40)_{\mathrm{stat}}(100)_{\phi}$\\
        32D & $66.05(48)_{\mathrm{stat}}(106)_{\phi}$\\
        32Df & $61.33(48)_{\mathrm{stat}}(93)_{\phi}$\\
        48I & $60.61(37)_{\mathrm{stat}}(99)_{\phi}$\\
        64I & $60.88(48)_{\mathrm{stat}}(97)_{\phi}$\\
        \hline
    \end{tabular}
    \caption{Results for $a_{\mu}^{\pi^0\mathrm{-pole}}$ for the gauge ensembles with near-physical pion masses. These results include finite-volume corrections, unphysical pion mass corrections and disconnected contributions. The uncertainty labeled with the subscript $\phi$ indicates that it stems from the five different parameterizations.}
    \label{tab:amu}
\end{table}

\begin{figure}[htbp!]
    \centering
    \includegraphics[width=0.48\textwidth]{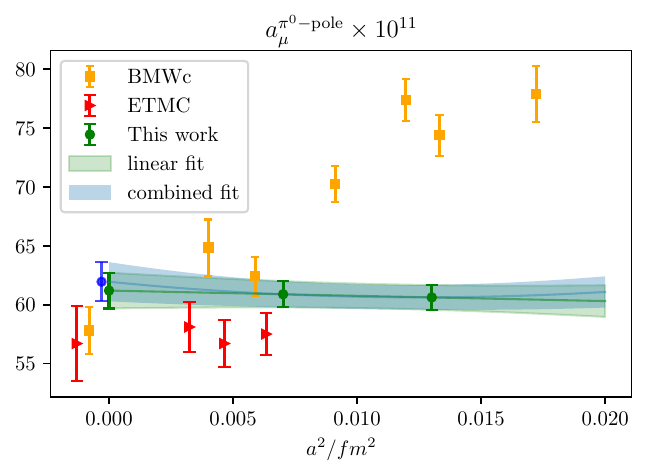}
    \caption{Lattice results for $a_{\mu}^{\pi^0\text{-pole}}$ as a function of the lattice spacing squared, $a^2$. 
    For each data point from this work, the uncertainty reflects both statistical error and systematic variations from different parameterizations of $\phi_\pi(x^2,u)$. These results are
    compared with previous findings from the BMWc~\cite{Gerardin:2023naa} 
    and ETMC~\cite{ExtendedTwistedMass:2023hin} Collaborations. In the continuum limit, our results are slightly higher than those from other calculations, 
    primarily because our disconnected contributions are positive, whereas in other studies, they contribute negatively.}
    \label{fig:result_lat}
\end{figure}

Table~\ref{tab:amu} lists $a_{\mu}^{\pi^0\mathrm{-pole}}$ results for ensembles with near-physical pion masses, incorporating finite-volume corrections, unphysical pion mass corrections, and disconnected contributions (see Supplemental Material~\cite{SM} for details). Before finite-volume corrections, the 24D and 32D ensembles—differing only by volume—exhibit clear finite-volume effects. After correction, the results differ by less than 1\,$\sigma$, likely due to statistical fluctuations; however, this difference is conservatively treated as a systematic uncertainty.
Using additional 32IcH ensemble with a heavier pion mass, we estimate and correct for pion mass dependence. These corrections are small relative to other systematic effects. The primary uncertainty stems from five different parameterizations. The CZ parameterization yields the maximum value for $a_{\mu}^{\pi^0\mathrm{-pole}}$, while the point-like yields the minimum. The central result is quoted as the average, with half of their difference as the systematic uncertainty.
The continuum extrapolations are shown in Fig.~\ref{fig:result_lat}. A linear $a^2$ fit to 48I and 64I data points gives $a_{\mu}^{\pi^0\mathrm{-pole}} = 61.2(1.1)_{\mathrm{stat}}(1.0)_{\phi}$. To estimate residual $O(a^4)$ lattice artifacts, we use a combined quadratic fit incorporating the 32D and 32Df data, yielding $a_{\mu}^{\pi^0\mathrm{-pole}}=61.9(1.7)_{\mathrm{stat}}(1.0)_{\phi}$. The difference between these fits serves as  the residual artifact estimate. Details of the quadratic fit are provided in the Supplemental Material~\cite{SM}.
The final result is
\be
a_{\mu}^{\pi^0\mathrm{-pole}} = 61.2(1.1)_{\mathrm{stat}}(1.0)_{\phi}(0.7)_{a}(0.4)_{\mathrm{FV}}\times 10^{-11}.
\ee
The model-independent component $a_{0,0}^{\pi^0\mathrm{-pole}}$ is given by $a_{0,0}^{\pi^0\mathrm{-pole}}=61.2(1.1)_{\mathrm{stat}}\times 10^{-11}$,
showing strong consistency as averaging the CZ and point-like parameterizations suppresses the contribution from the $\varphi_2(r)$ term.

We also calculate the $\pi^0$ decay width as $\Gamma_{\pi^0\to\gamma\gamma}=7.60(27)$ eV, which is $\sim 0.7\,\sigma$ below the experimental result of $\Gamma_{\pi^0\to\gamma\gamma}=7.80(12)$ eV~\cite{PrimEx-II:2020jwd}.

\section{Conclusion}

We have developed a method to calculate $a_{\mu}^{\pi^0\text{-pole}}$, the $\pi^0$-pole contribution to the HLbL in the muon $g-2$. Our approach includes an $SO(4)$ average to compute the TFF at arbitrary space-like photon momenta and introduces a structure function, $\phi_\pi(x^2,u)$, to mitigate the signal-to-noise problem. As demonstrated in the Supplemental Material, $\phi_\pi(x^2,u) = 0$ for $u \notin [0,1]$, providing an additional constraint that improves the lattice results compared to other methods. To estimate the systematic effects associated with $\phi_\pi(x^2,u)$, we employ a Gegenbauer expansion and introduce five parameterizations of the structure function, 
spanning a wide range of large-momentum behaviors for the TFF.

\begin{figure}[htbp!]
    \centering
    \includegraphics[width=0.48\textwidth]{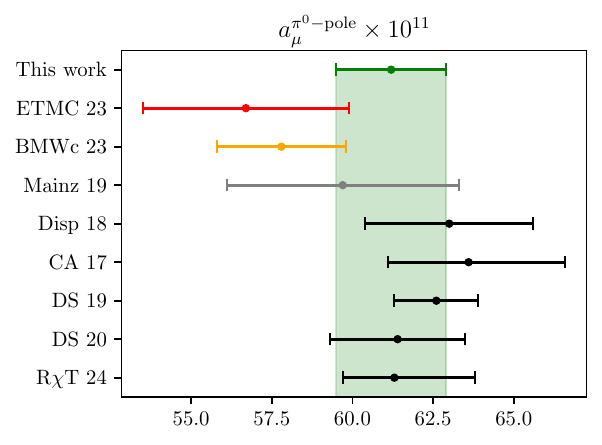}
    \caption{Comparison of various results for $a_{\mu}^{\pi^0\text{-pole}}$, including lattice calculations, dispersive analyses, and phenomenological studies. 
    The results, listed from top to bottom, include: lattice QCD calculations from ETMC 23~\cite{ExtendedTwistedMass:2023hin}, BMWc 23~\cite{Gerardin:2023naa}, 
    and Mainz 19~\cite{Gerardin:2019vio}; dispersive analysis (Disp 18)~\cite{Hoferichter:2018dmo}; Canterbury approximants (CA 19)~\cite{Masjuan:2017tvw}; 
    Dyson-Schwinger approach (DS 19)~\cite{Eichmann:2019tjk} and (DS 20)~\cite{Raya:2019dnh}; and
    resonance chiral theory (R$\chi$T 24)~\cite{Estrada:2024cfy}.}
    \label{fig:comparison}
\end{figure}

In this calculation, we conduct a detailed analysis of various systematic effects. 
In Fig.~\ref{fig:comparison}, we compare our final result with other lattice QCD calculations\cite{ExtendedTwistedMass:2023hin,Gerardin:2023naa,Gerardin:2019vio}, dispersive analyses~\cite{Hoferichter:2018dmo}, and phenomenological studies~\cite{Masjuan:2017tvw,Eichmann:2019tjk,Raya:2019dnh,Estrada:2024cfy}. 
Despite methodological differences, our results align well with other lattice calculations but yield a larger value due to the opposite sign we obtain for disconnected contributions, bringing our lattice results closer to dispersive determinations.
The uncertainties of our final result are comparable to those of previous calculations. In our approach, the statistical uncertainties are significantly reduced due to the application of the new method. As shown in Table~\ref{tab:amu}, for each ensemble, the dominant uncertainties now stem from the five parameterizations of the structure function. 
Future work can reduce systematic uncertainties further by calculating the hadronic function $H(x^2,P \cdot x)$ at more values of $P \cdot x$ by boosting the pion.

Expanding on methods used for decay amplitudes in processes like $\pi^0 \to e^+e^-$ and $\eta_c \to \gamma\gamma$, our approach holds promise for future calculations of the $\eta$- and $\eta'$-pole contributions to the HLbL in the muon $g-2$, where heavier meson masses allow for a wider range of $P \cdot x$ in $H(x^2, P \cdot x)$.

\begin{acknowledgments}
{\bf Acknowledgments} -- X.F., T.L., C.L. and Q.Y.L. were supported in part by NSFC of China under Grant No. 12125501, No. 12293060, No. 12293063, and No. 12141501,
and National Key Research and Development Program of China under No. 2020YFA0406400.
L.C.J. acknowledges support by DOE Office of Science Early Career Award No. DE-SC0021147 and DOE Award No. DE-SC0010339.
M.B. is (partially) supported by ICSC - Centro Nazionale di Ricerca in High Performance Computing, Big Data and Quantum Computing, funded by European Union – NextGenerationEU.
The research reported in this work was carried out using the computing facilities at Chinese National Supercomputer Center in Tianjin.
It also made use of computing and long-term storage facilities of the USQCD Collaboration, which are funded by the Office of Science of the U.S. Department of Energy.
An award of computer time was provided by the ASCR Leadership Computing Challenge (ALCC) and Innovative and Novel Computational Impact on Theory and Experiment (INCITE) programs.  Data used in this work was generated in part using CPS~\cite{Jung:2014ata},  GPT~\cite{GPT}, Grid~\cite{Boyle:2016lbp,Yamaguchi:2022feu}, and QLattice~\cite{QLattice}.
\end{acknowledgments}

\appendix




\bibliography{pi-pole}

\clearpage

\setcounter{page}{1}
\renewcommand{\thepage}{Supplemental Material -- S\arabic{page}}
\setcounter{table}{0}
\renewcommand{\thetable}{S\,\Roman{table}}
\setcounter{equation}{0}
\renewcommand{\theequation}{S\,\arabic{equation}}
\setcounter{figure}{0}
\renewcommand{\thefigure}{S\,\arabic{figure}}

\section{Supplemental Material}

\subsection{Signal-to-noise problem}

\begin{figure}[htb]
\centering
\includegraphics[width=0.48\textwidth,angle=0]{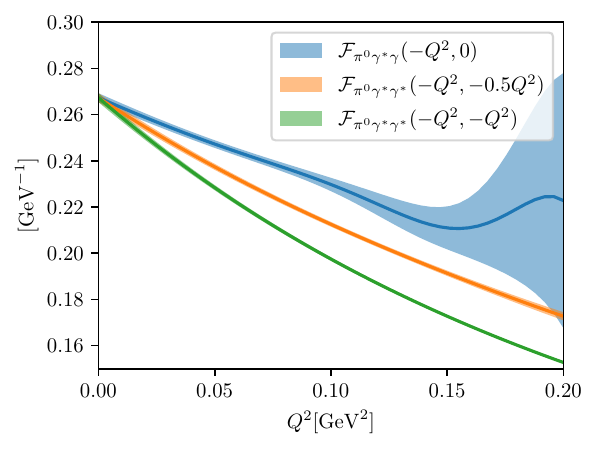}
\caption{
Numerical results of $\mathcal{F}_{\pi^0\gamma^*\gamma}(-Q^2,0)$, $\mathcal{F}_{\pi^0\gamma^*\gamma^*}(-Q^2,-0.5\,Q^2)$ and $\mathcal{F}_{\pi^0\gamma^*\gamma^*}(-Q^2,-{Q}^2)$
as functions of $Q^2$. These results are calculated using Eq.~(\ref{eq:TFF_SO3}), with ensemble 64I serving as an example.
}
\label{fig:TFF_SO3}
\end{figure}

In this section, we address the signal-to-noise problem that arises in the integral with the $SO(3)$-averaged weight function. By substituting
Eq.~(\ref{eq:weight_function_SO3})
into Eq.~(\ref{eq:relation}), we obtain
\ba
\label{eq:TFF_SO3}
\mathcal{F}_{\pi^0\gamma^*\gamma^*}(-Q^2,-{Q'}^2) &=& \int d^4x \,e^{(E-\frac{1}{2}m_{\pi})t}\frac{|\vec{x}|}{|\vec{q}|}j_1(|\vec{q}||\vec{x}|) 
\nn\\
&&\hspace{1cm}\times H(x^2,P\cdot x)
\ea
with
\be
P=(im_\pi,\vec{0}),\quad Q=(iE,\vec{q}),\quad Q'=(i(m_\pi-E),-\vec{q}).
\nn
\ee
To describe the virtuality of the photons, we introduce two parameters, $\rho$ and $\sigma$
\be
\rho m_\pi^2=-Q^2=E^2-|\vec{q}|^2,\quad \sigma m_\pi^2=-{Q'}^2=(m_\pi-E)^2-|\vec{q}|^2.
\nn
\ee
The quantities $E$ and $|\vec{q}|$ can then be expressed in terms of $\rho$ and $\sigma$ as
\be
E=\frac{1+\rho-\sigma}{2}m_\pi,\quad |\vec{q}|=\frac{\sqrt{(\rho-\sigma)^2-2\rho-2\sigma+1}}{2}m_\pi.
\nn
\ee
When $|\rho-\sigma|\gg 1$, or equivalently $|Q^2-{Q'}^2|\gg m_\pi^2$, we find that $|E|\gg m_\pi$. In this scenario, we expect the TFF to be highly noisy.
A specific example of this is when $Q^2\gg m_\pi^2$ and ${Q'}^2=0$. On the other hand, when $\rho=\sigma$, or equivalently $Q^2={Q'}^2$, we have $E=\frac{m_\pi}{2}$. 
In this case, there is no signal-to-noise problem.
In Fig.~\ref{fig:TFF_SO3}, we present the lattice results of the TFF for three cases: ${Q'}^2/Q^2=0$, $0.5$ and $1$. These results confirm our expectations.

\subsection{On-shell transition form factor}

\begin{table}[htbp!]
    \centering
    \begin{tabular}{l| l l c c| c c}
        \hline
    & \multicolumn{5}{c}{$\mathcal{F}_{\pi^0\gamma\gamma}(0,0)\times10^{3}~[\mathrm{GeV}^{-1}]$ using $SO(3)$ average} \\
        \hline
        Ens &  conn. diag. & FV corr. & mass corr. & disc. diag. & total \\
        \hline
        24D & $283.0(1.6)$ & 4.5(3) & 0.60(5) & 4.1(7) & 291.5(1.7) \\
        32D & $288.8(2.2)$ &  0.44(5) &  0.60(5) & - &  293.3(2.2) \\
        32Df & $270.8(2.6)$ & 4.8(7) &  0.63(5) & - & 279.6(2.7) \\
        48I & $268.2(1.9)$ & 1.15(9) & 0.37(3)  & - & 273.1(1.9) \\
        64I &  $267.3(1.9)$ & 1.3(1)  & 0.31(3)  & 3.4(3) & 272.3(1.9) \\
        24DH & $249.1(0.9)$ & 0.90(5) & - & - & - \\ 
        24IH & $209.3(0.9)$ & 26.2(8) & - & - & - \\
        32IcH & $233.5(1.2)$ & 7.3(3) & - & - & - \\
        \hline
        \hline
        & \multicolumn{5}{c}{$\mathcal{F}_{\pi^0\gamma\gamma}(0,0)\times10^{3}~[\mathrm{GeV}^{-1}]$ using $SO(4)$ average} \\
        \hline
        Ens &  conn. diag. & FV corr. & mass corr. & disc. diag. & total \\
        \hline
        24D & $284.6(1.7)$ & 4.3(3) & 0.60(5) &  4.1(7) & 292.9(1.7)  \\
        32D & $289.7(2.2)$ & 0.42(4) & 0.60(5) & - & 294.1(2.3) \\
        32Df & $272.9(2.7)$ & 4.6(7) & 0.62(5) & - & 281.5(2.8) \\
        48I & $268.6(1.9)$ &  1.09(9) & 0.36(3) & - & 273.5(1.9) \\
        64I & $267.7(1.9)$ & 1.3(1)  & 0.31(3)  & 3.4(3) & 272.7(1.9) \\
        24DH & $250.0(0.8)$ & 1.00(5) & - & - & - \\
        24IH & $210.2(0.9)$ & 25.8(8) & - & - & - \\
        32IcH & $234.0(1.2)$ & 7.4(3) & - & - & - \\
        \hline
    \end{tabular}
    \caption{Numerical results of $\mathcal{F}_{\pi^0\gamma\gamma}(0,0)$ across various ensembles calculated using both the $SO(3)$ averaging method~(\ref{eq:TFF_SO3}) and $SO(4)$ averaging method~(\ref{eq:master_formula}). For the latter, we use $\phi_\pi(x^2,u)=1$ (VMD parameterization) as input. 
Different parameterizations for $\phi_\pi(x^2, u)$ have a negligible effect on the determination of $\mathcal{F}_{\pi^0\gamma\gamma}(0,0)$. Finite-volume effects, unphysical pion mass effects, and contributions from disconnected diagrams are also listed.}
    \label{tab:results_F00}
\end{table}

In Table~\ref{tab:results_F00}, we present the lattice results for the on-shell TFF $\mathcal{F}_{\pi^0\gamma\gamma}(0,0)$ across various ensembles. For calculating $\mathcal{F}_{\pi^0\gamma\gamma}(0,0)$, there is no signal-to-noise problem, allowing both the $SO(3)$ and $SO(4)$ averaging methods to yield results with comparable precision.
However, compared to the $SO(3)$ results in the upper part of the table, the $SO(4)$ results in the lower part exhibit systematic upward shifts. These shifts are primarily due to lattice artifacts, which tend to be larger at coarser lattice spacings and diminish as the lattice spacing becomes finer.
We also compute the finite-volume effects, the unphysical pion mass effects, and the contributions from disconnected diagrams, as shown in Table~\ref{tab:results_F00}. 
For the disconnected diagrams, we calculate contributions for the 24D and 64I ensembles, finding good consistency between them; thus, we use the 64I results as corrections from the disconnected diagrams. 
Regarding finite-volume effects, these can reach 4.3-4.8$\times10^{-3}$ GeV$^{-1}$ for the two ensembles with lattice sizes of $L \approx 4.6$ fm, specifically 32Df and 24D. 
As the lattice size increases to $L \approx 5.5$ fm (48I and 64I ensembles), the finite-size correction reduces to 1.1-1.3$\times10^{-3}$ GeV$^{-1}$. 
For the 32D ensemble with $L \approx 6.2$ fm, the finite-size effect is further suppressed to 0.4$\times10^{-3}$ GeV$^{-1}$.
In contrast, the 24IH and 32IcH ensembles, which have heavier pion masses, exhibit relatively large finite-volume effects due to their smaller lattice sizes ($L=2.7$ and 3.5 fm, respectively). However, after finite-volume corrections, results from these two ensembles align more closely, with small differences likely from residual effects in 24IH. We thus use only 32IcH for mass correction. The 24DH ensemble shows larger lattice artifacts due to its coarse spacing and is therefore excluded from mass correction. Overall, for ensembles with near-physical pion mass, mass corrections remain relatively small.
Further discussion of these systematic effects can be found in later sections.

\begin{table}[htbp!]
    \centering
    \begin{tabular}{|c|c|c|}
    \hline
    \multicolumn{3}{|c|}{$\mathcal{F}_{\pi^0\gamma\gamma}(0,0)\times10^{3}~[\mathrm{GeV}^{-1}]$}\\
    \hline
    Method & linear fit &  quadratic fit \\
    \hline
    SO(3) total & 271.4(4.7)  & 272.5(5.4) \\
    SO(4) total & 271.8(4.7)  & 272.5(5.4) \\
    \hline
    \end{tabular}
    \caption{Continuum-extrapolated results for $\mathcal{F}_{\pi^0\gamma\gamma}(0,0)$ obtained from linear and combined quadratic fits.}
    \label{tab:results_F00_continuum}
\end{table}    

\begin{figure}[htb]
\centering
\includegraphics[width=0.48\textwidth,angle=0]{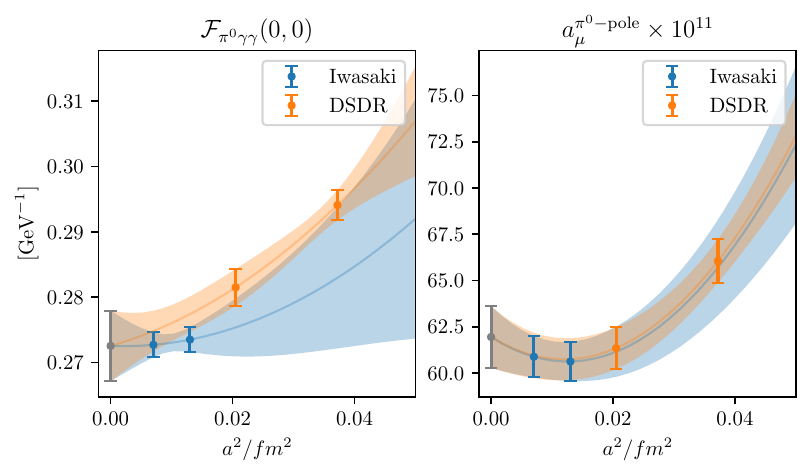}
\caption{
The combined quadratic fits for $\mathcal{F}_{\pi^0\gamma\gamma}(0,0)$ (left) and
$a_{\mu}^{\pi^0\mathrm{-pole}}$ (right).
}
\label{fig:quadratic_fit}
\end{figure}

Combining all contributions, we present the total results in the final column of Table~\ref{tab:results_F00}.
Since the DSDR ensembles with near-physical pion masses have relatively large lattice spacings, we use two Iwasaki ensembles, 48I and 64I, to perform a linear fit in $a^2$, obtaining the form factor in the continuum limit. To estimate the residual $O(a^4)$ lattice artifact, we carry out a combined fit across results from the 32D, 32Df, 48I, and 64I ensembles using the following functional form
\ba
\mathcal{F}_{\pi^0\gamma\gamma}(0,0)\Big|_{\mathrm{Iwasaki}}&=&\mathcal{F}_{\pi^0\gamma\gamma}(0,0)(1+c_1a^2+c_2a^4),
\nn\\
\mathcal{F}_{\pi^0\gamma\gamma}(0,0)\Big|_{\mathrm{DSDR}}&=&\mathcal{F}_{\pi^0\gamma\gamma}(0,0)(1+c_1'a^2+c_2a^4).
\ea
Since the Iwasaki and DSDR ensembles have different discretizations, we introduce separate parameters, $c_1$ and $c_1'$, for the linear term, while using a common parameter $c_2$ for the $a^4$ term across both discretizations. We interpret the difference between the linear and combined quadratic fits as the residual $O(a^4)$ systematic uncertainty. The fitting curves are displayed on the left side of Fig.~\ref{fig:quadratic_fit} and
the fit results for $\mathcal{F}_{\pi^0\gamma\gamma}(0,0)$ are 
summarized in Table~\ref{tab:results_F00_continuum}. We report the final results for $\mathcal{F}_{\pi^0\gamma\gamma}(0,0)$ as follows
\be
\mathcal{F}_{\pi^0\gamma\gamma}(0,0)=\begin{cases}
0.2714(47)_{\mathrm{stat}}(11)_{\mathrm{syst}}\mbox{ GeV}^{-1}, & \mbox{$SO(3)$}\\
0.2718(47)_{\mathrm{stat}}(7)_{\mathrm{syst}}\mbox{ GeV}^{-1}, & \mbox{$SO(4)$}
\end{cases},
\ee
which are 0.7-0.8\,$\sigma$ below the experimental value of $\mathcal{F}_{\pi^0\gamma\gamma}(0,0) = 0.2754(21)$ GeV$^{-1}$, 
obtained using the PrimEx-II decay width of $\Gamma_{\pi^0\to\gamma\gamma}=7.80(12)$ eV as input~\cite{PrimEx-II:2020jwd}. 
The corresponding TFF results imply a decay width of $\Gamma_{\pi^0\to\gamma\gamma}=7.58(26)_{\mathrm{stat}}(6)_{\mathrm{syst}}$ eV ($SO(3)$)
and $\Gamma_{\pi^0\to\gamma\gamma}=7.60(26)_{\mathrm{stat}}(4)_{\mathrm{syst}}$ eV ($SO(4)$), which are also $\sim0.8\,\sigma$ below
the experimental value. To maintain consistency throughout the paper, we use the $SO(4)$ result as the final value for the decay width.

The combined quadratic fit for $a_{\mu}^{\pi^0\mathrm{-pole}}$ is performed in the same manner as for $\mathcal{F}_{\pi^0\gamma\gamma}(0,0)$. The fitting curves are shown on the right side of Fig.~\ref{fig:quadratic_fit}, and the fit results are presented in the main text.

\subsection{Derivation of Eq.~(\ref{eq:structure_func})}

To describe the $P\cdot x$ dependence in $H(x^2,P\cdot x)$ , we express it through a Fourier transformation
\be
H(x^2,P\cdot x)=\int_{-\infty}^\infty du\,e^{i\left(u-\frac{1}{2}\right)P\cdot x}\tilde{H}(x^2,u).
\ee
By defining the pion structure function $\phi_\pi$ as
\be
\phi_\pi(x^2,u)\equiv\frac{\tilde{H}(x^2,u)}{H(x^2,0)},
\ee
we immediately obtain
\be
H(x^2,P\cdot x)=\int_{-\infty}^\infty du\,e^{i\left(u-\frac{1}{2}\right)P\cdot x}\phi_\pi(x^2,u)H(x^2,0).
\ee
Setting $P\cdot x=0$ leads to a normalization condition for $\phi_\pi$
\be
\int_{-\infty}^\infty du\,\phi_\pi(x^2,u)=1, \quad \forall x^2.
\ee

Next, we will prove $\phi_{\pi}(x^2,u) = 0$ when $u\notin[0,1]$.
If it holds, we recover Eq.~(\ref{eq:structure_func}) from the main paper.
To do this, we define an operator $\bar{O}_\pi$ that creates only the pion state from vacuum and is normalized as
\be
\langle \pi(P)|\bar{O}_\pi(y)|0\rangle=e^{-iP\cdot y}.
\ee
Using $\bar{O}_\pi$, we construct the three-point correlation function as
\be
C_{\mu\nu}(x,y) =\left\langle 0\left| T\left[ J_\mu\left(\frac{x}{2}\right)J_\nu\left(-\frac{x}{2}\right) \bar{O}_\pi(y)\right]\right| 0 \right\rangle.
\ee
We choose the temporal component of $y$ to satisfy $y_0<-\frac{1}{2}|x_0|$ and insert the pion state into the three-point function
\ba
    C_{\mu\nu}(x,y) &= &  \int \frac{d^3\vec{p}}{(2\pi)^3} \varepsilon_{\mu\nu\alpha\beta}x_\alpha P_\beta H(x^2,P\cdot x)\frac{1}{2E_{\pi}}e^{-iP\cdot y} 
    \nn\\
    &= &\int \frac{d^3\vec{p}}{(2\pi)^3} \varepsilon_{\mu\nu\alpha\beta}x_\alpha \left(i\frac{\partial}{\partial y_\beta}\right)
    \nn\\
    &&\hspace{0.5cm} \int_{-\infty}^{\infty} du \,e^{i\left(u-\frac{1}{2}\right)P\cdot x} \tilde{H}(x^2,u) \frac{1}{2E_{\pi}}e^{-iP\cdot y}
\nn\\
    &= &\varepsilon_{\mu\nu\alpha\beta}x_\alpha \left(i\frac{\partial}{\partial y_\beta}\right)
    \nn\\
   &&\hspace{0.5cm} \int_{-\infty}^{\infty} du \, \tilde{H}(x^2,u)  G\left(\left(u-\frac{1}{2}\right)x-y\right).
\ea
In the last line, we introduce the pion propagator
\be
G(x)=\int\frac{d^4P}{(2\pi)^4}\frac{e^{iP\cdot x}}{P^2+m_\pi^2},
\ee
which becomes singular as $x\to0$.

Next, we draw a straight line through $-\frac{x}{2}$ and $\frac{x}{2}$, placing $y$ on this line. For any $u\notin [0,1]$,
it is always possible to choose $y$ such that it approaches $\left(u-\frac{1}{2}\right)x$, causing the pion propagator to become singular.
On the other hand, there is no singularity associated with the correlation function $C_{\mu\nu}(x,y)$ when $y_0<-\frac{1}{2}|x_0|$. 
This leads to a contradiction, implying that $\tilde{H}(x^2,u) = 0$, $\forall u\notin [0,1]$, and therefore, the same holds for $\phi_\pi(x^2,u)$.

\subsection{Influence from the structure function $\phi_\pi(x^2,u)$}

In this section, we discuss the influence of the structure function $\phi_\pi(x^2,u)$ on the final determination of $a_{\mu}^{\pi^0\mathrm{-pole}}$.
We begin by examining the application of Eq.~(\ref{eq:relation_SO4}). Under Gegenbauer expansion~(\ref{eq:Gegenbauer}), the TFF can be expressed as
\ba
\label{eq:TFF_f2n}
&&\mathcal{F}_{\pi^0\gamma^*\gamma^*}(-Q^2,-Q'^2)
=12 \int d^4x \,H(x^2,0)
\nn\\
&&\hspace{2cm}\times\sum_{n=0}^\infty \varphi_{2n}(r) f_{2n}(Q^2,{Q'}^2,r),
\ea
with
\be
\label{eq:f2n}
f_{2n}(Q^2,{Q'}^2,r)\equiv\int_0^1 du\,\frac{J_2(|\bar{K}|r)}{|\bar{K}|^2}u(1-u)C_{2n}^{(\frac{3}{2})}(2u-1).
\ee

Given a specific value of $r$, we calculate $f_{2n}(Q^2,{Q'}^2,r)$ at the following points
\be
\{Q^2,{Q'}^2\}=\{Q_1^2,0\},\,\{Q_2^2,Q_3^2\},\,\{Q_3^2,0\},\,\{Q_1^2,Q_2^2\}
\ee
using the one-dimensional integral in Eq.~(\ref{eq:f2n}).
With $f_{2n}(Q^2,{Q'}^2,r)$ as inputs, we substitute~(\ref{eq:TFF_f2n}) into Eq.~(\ref{eq:amu-pion-pole}) and numerically compute the integral $\int dQ_1\int dQ_2\int d\tau$.
This yields
\ba
\label{eq:pion_pole_part_data}
a_{\mu}^{\pi^0\mathrm{-pole}}&=&\int d^4x_1\int d^4x_2\,H(x_1^2,0)H(x_2^2,0)
\nn\\
&&\times\sum_{n\ge m} \varphi_{2n}(r_1)\varphi_{2m}(r_2) \rho_{2n,2m}(r_1,r_2),
\ea
where the weight function $\rho_{2n,2m}(r_1,r_2)$ is determined numerically.

Due to the lattice discretization, the spacetime integrals $\int d^4x_1\int d^4x_2$ are replaced by summations
\be
\int d^4x_1\int d^4x_2 \quad\rightarrow\quad \sum_{r_1,r_2}\sum_{x_1,x_2}\delta_{|x_1|,r_1}\delta_{|x_2|,r_2}.
\ee 
This implies that we need to compute $\rho_{2n,2m}(r_1,r_2)$ at numerous points of $\{r_1,r_2\}$, totaling $L^2\times L^2$ points. 
To reduce computational cost, we use interpolation. For example, along the $r_1$ direction, if $r_1 \leq 0.5$ fm, we calculate $\rho_{2n,2m}(r_1, r_2)$ at every point of $r_1$. 
For $r_1 > 0.5$ fm, we calculate it with a spacing of 0.05 fm and then perform numerical interpolation to obtain $\rho_{2n,2m}(r_1, r_2)$ at all points of $r_1$. 
The same treatment is applied to $r_2$.

\begin{figure}[htb]
\centering
\includegraphics[width=0.48\textwidth,angle=0]{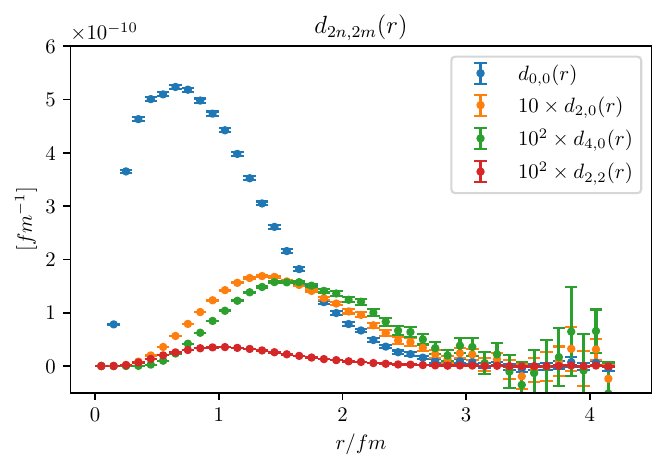}
\caption{
The coefficients $d_{2n,2m}(r)$ as functions of $r$. There coefficients are calculated using the 64I ensemble. The determination of $d_{0,0}(r)$, $d_{2,0}(r)$, and $d_{4,0}(r)$ does not
require the input from the structure function. For $d_{2,2}(r)$, we have used the CZ parameterization, which provides the largest $|\varphi_2|$ among the five parameterizations considered in this study. The other curves are model independent.
}
\label{fig:d2n2m}
\end{figure}

Using the spacetime summation, we obtain
\ba
a_{\mu}^{\pi^0\mathrm{-pole}}&=&\sum_{r_1,r_2}\sum_{n\ge m}c_{2n,2m}(r_1,r_2)\varphi_{2n}(r_1)\varphi_{2m}(r_2)
\nn\\
&=&a_{0,0}^{\pi^0\mathrm{-pole}}+a_{2,0}^{\pi^0\mathrm{-pole}}+a_{2,2}^{\pi^0\mathrm{-pole}}+a_{4,0}^{\pi^0\mathrm{-pole}}+\cdots
\nn\\
\ea
with
\ba
c_{2n,2m}(r_1,r_2)&=&\sum_{x_1,x_2}\delta_{|x_1|,r_1}\delta_{|x_2|,r_2}\rho_{2n,2m}(r_1,r_2)
\nn\\
&&\hspace{1cm}\times H(x_1^2,0)H(x_2^2,0).
\ea
We define 
\be
a_{2n,2m}^{\pi^0\mathrm{-pole}}=\sum_{r}d_{2n,2m}(r)\varphi_{2n}(r)
\ee
and
\be
d_{2n,2m}(r)=\sum_{{r'}}c_{2n,2m}(r,r')\varphi_{2m}(r').
\ee
For $d_{2n,0}$, it can be calculated in a model-independent manner. However, for $d_{2,2}$, we need to use $\varphi_{2}(r)$ from various parameterizations~(\ref{eq:model}) as input.
The results of $d_{0,0}$, $d_{2,0}$, $d_{4,0}$ and $d_{2,2}$ as functions of $r$ are displayed in Fig.~\ref{fig:d2n2m}. In comparison to $d_{0,0}(r)$, $d_{2,0}(r)$ reaches a few percent, while $d_{4,0}(r)$ and $d_{2,2}(r)$ are further suppressed by an additional factor of 10. 

\begin{figure}[htb]
\centering
\includegraphics[width=0.48\textwidth,angle=0]{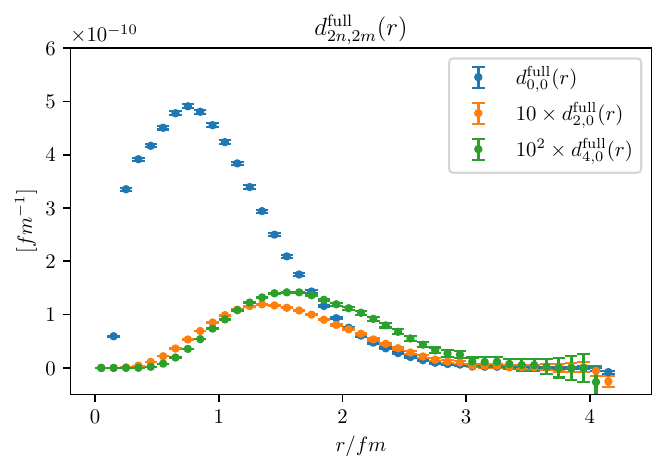}
\caption{
The coefficients $d_{0,0}^{\mathrm{full}}(r)$, $d_{2,0}^{\mathrm{full}}(r)$ and $d_{4,0}^{\mathrm{full}}(r)$ as functions of $r$.
}
\label{fig:d2n2m_full}
\end{figure}

Next, we explore the application of Eq.~(\ref{eq:master_formula}). We can express $H(x^2,0)$ as
\ba
H(x^2,0)&=&\frac{H(x^2,P\cdot x)}{\int du\,e^{i\left(u-\frac{1}{2}\right)P\cdot x}\phi_\pi(x^2,u)}
\nn\\
&=&\frac{H(x^2,P\cdot x)}{\sum_{n=0}^\infty \varphi_{2n}(r)g_{2n}(P\cdot x)},
\ea
with $g_{2n}(P\cdot x)$ representing a series of analytically known functions. For $n>0$, the $\varphi_{2n}(r)$ term can be treated as a correction, leading to
\be
\label{eq:relation_expansion}
H(x^2,0)=\frac{H(x^2,P\cdot x)}{g_0(P\cdot x)}\left(1-\sum_{n>0}\frac{\varphi_{2n}(r)g_{2n}(P\cdot x)}{g_0(P\cdot x)}\right)+\cdots
\ee 
By substituting Eq.~(\ref{eq:relation_expansion}) into Eq.~(\ref{eq:pion_pole_part_data}) and isolating the terms that depend linearly on $\varphi_{2n}(r_1)\varphi_{2m}(r_2)$, we obtain
$a_{2n,2m}^{\pi^0\mathrm{-pole}}=\sum_{r}d_{2n,2m}^{\mathrm{full}}(r)\varphi_{2m}(r)$. 
We use the superscript ``full'' to indicate that $d_{2n,2m}^{\mathrm{full}}(r)$ is computed using the full dataset of $H(x^2,P\cdot x)$, rather than just $H(x^2,0)$. 
The coefficients $d_{2n,2m}^{\mathrm{full}}(r)$ as functions of $r$ are shown in Fig.~\ref{fig:d2n2m_full}. 

\subsection{Finite-volume effects}
At long distances, the hadronic function $\mathcal{H}_{\mu\nu}(x)$ is assumed to be dominated by contributions from a vector meson $V$ and $\pi V$ states.
For simplicity, the vector mesons in both the isovector and isoscalar channels are taken to have the same mass.
 For $t>0$, we have
\ba
&&\mathcal{H}_{\mu\nu}^{\mathrm{LD}}(x)=\sum_\lambda \int \frac{d^3\vec{q}}{(2\pi)^3}\times
\nn\\
&&\left[\langle 0|J_\mu|V(Q_{\mathrm{on}},\lambda)\rangle \frac{e^{-(E_V-\frac{1}{2}m_\pi)t+i\vec{q}\cdot\vec{x}}}{2E_V}
\langle V(Q_{\mathrm{on}},\lambda)|J_\nu|\pi(P)\rangle+\right.
\nn\\
&&\left.\langle 0|J_\mu|\pi(P)V(Q_{\mathrm{on}},\lambda)\rangle \frac{e^{-(E_V+\frac{1}{2}m_\pi)t+i\vec{q}\cdot\vec{x}}}{2E_V}
\langle V(Q_{\mathrm{on}},\lambda)|J_\nu|0\rangle\right],
\nn\\
\ea
where $Q_{\mathrm{on}}=(iE_V,\vec{q})$ and $E_V$ is the energy of the vector meson. The index $\lambda$ denotes the polarization direction of 
the vector meson. The matrix elements are decomposed as
\ba
&&\langle0|J_\mu|V(Q,\lambda)\rangle=m_V f_V\epsilon_{\mu}(Q,\lambda),
\nn\\
&&\langle V(Q,\lambda)|J_\nu|0\rangle=m_V f_V\epsilon_{\nu}^*(Q,\lambda),
\nn\\
&&\langle V(Q,\lambda)|J_\nu|\pi(P)\rangle=i\,\varepsilon_{\nu\mu\alpha\beta}\epsilon^{*\mu}(Q,\lambda)Q^\alpha P^\beta F_{\pi V}(P,Q),
\nn\\
&&\langle0|J_\mu| \pi(P)V(Q,\lambda)\rangle=-i\,\varepsilon_{\mu\nu\alpha\beta}\epsilon^\nu(Q,\lambda) Q^\alpha P^\beta F_{\pi V}(P,-Q),
\nn\\
\ea
assuming that the vector meson is treated as a stable particle and that rescattering effects between $\pi$ and $V$ are neglected.
Here, $f_V$ represents the vector meson decay constant, and $F_{\pi V}(P,Q)$ is a form factor dominated by the vector meson pole
\be
F_{\pi V}(P,Q)=\frac{f_{\pi V}}{(P-Q)^2+m_V^2}.
\ee
A similar procedure applies for $t<0$. Using the polarization sum
$\sum_\lambda\epsilon_{\mu}(Q,\lambda)\epsilon_{\nu}^*(Q,\lambda)=\frac{Q_\mu Q_\nu}{-m_V^2}-g_{\mu\nu}$, we derive
\ba
\label{eq:VMD}
&&\mathcal{H}_{\mu\nu}^{\mathrm{LD}}(x)=i\int \frac{d^3\vec{q}}{(2\pi)^3}\frac{f_{\pi VV}\varepsilon_{\mu\nu\alpha\beta}Q_{\mathrm{on}}^\alpha P^\beta}{2E_V m_\pi}\times
\nn\\
&&\hspace{0.8cm}\left[\frac{e^{-(E_V-\frac{1}{2}m_\pi)|t|+i\vec{q}\cdot\vec{x}}}{2E_V-m_\pi}-\frac{e^{-(E_V+\frac{1}{2}m_\pi)|t|+i\vec{q}\cdot\vec{x}}}{2E_V+m_\pi}\right],
\ea
with $f_{\pi VV}\equiv m_V f_V f_{\pi V}$. 
In infinite volume, we will later demonstrate that $\mathcal{H}_{\mu\nu}^{\mathrm{LD}}(x)$, as defined in Eq.~(\ref{eq:VMD}), produces a structure function $\phi_\pi^{LD}(x^2,u)\approx 1$.
In a finite volume, the hadronic function $\mathcal{H}_{\mu\nu}^{\mathrm{LD}}(L,x)$ is obtained by replacing $\int \frac{d^3\vec{q}}{(2\pi)^3}$ with $\frac{1}{L^3}\sum_{\vec{q}}$ in Eq.~(\ref{eq:VMD}). 
The scalar function $H^{\mathrm{LD}}(L,x^2,P\cdot x)$ is then defined as
\be
\label{eq:H_LD}
H^{\mathrm{LD}}(L,x^2,P\cdot x)=-
\frac{\varepsilon_{\mu\nu\alpha\beta}x^\alpha P^\beta \mathcal{H}_{\mu\nu}^{\mathrm{LD}}(L,x)}{2(x^2 P^2-(P\cdot x)^2)}.
\ee

\begin{table}
    \begin{tabular}{|c|c|c|c|c|}
        \hline
        Ensemble & $m_V$ [GeV] & $f_{\pi VV}$ [GeV$^3$] & $\tilde{\alpha}_{\mathrm{VMD}}$ & $\chi^2$/d.o.f  \\
        \hline
        24D & 0.719(10)  & 0.072(6) & 0.02566(63) & 0.7 \\
        32D & 0.713(10)  & 0.069(5) & 0.02533(58) & 0.9 \\
        32Df & 0.705(18)  & 0.059(8) & 0.02198(65) & 0.5\\
        48I & 0.716(9)  & 0.062(4) & 0.02207(43) & 0.2 \\
        64I & 0.716(12)  & 0.063(5) & 0.02214(51) & 0.6 \\
        24DH & 0.843(7)  & 0.119(5) & 0.02597(35) & 0.3 \\
        24IH & 0.853(7)  & 0.113(3) & 0.02478(18)  & 0.6 \\
        32IcH & 0.839(6)  & 0.118(4) & 0.02395(19)  & 0.5 \\
        \hline
        \end{tabular}
        \caption{The fitting parameters $m_V$, $f_{\pi VV}$, and the ratio $\tilde{\alpha}_{\mathrm{VMD}}=\frac{F_\pi f_{\pi VV}}{m_V^4}$ for all ensembles.}
        \label{tab:mrho_frhopipi}
\end{table}

\begin{figure}[htbp!]
    \centering
    \includegraphics[width=0.48\textwidth]{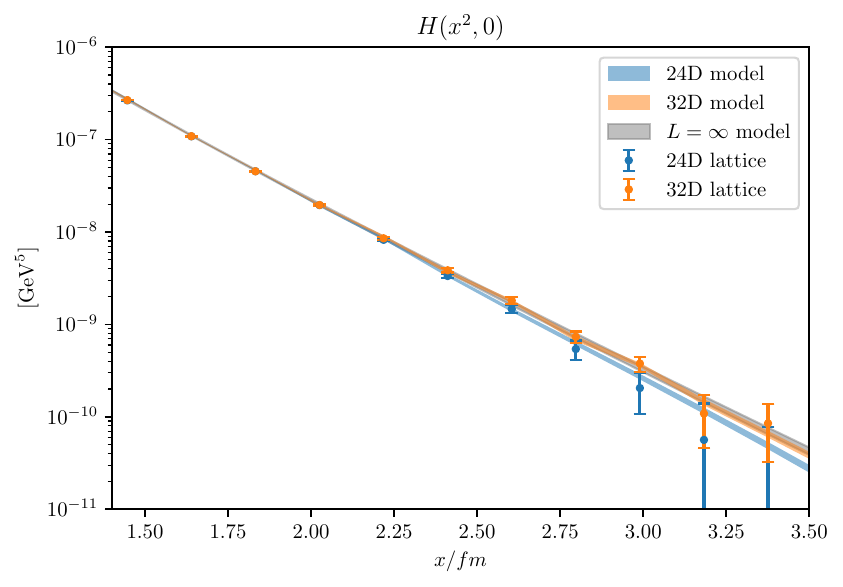}
    \caption{Fit of $H^{\mathrm{LD}}(L,x^2,0)$ to lattice data for the 24D and 32D ensembles.}
    \label{fig:VMD}
\end{figure}

In the master formula~(\ref{eq:master_formula}), the lattice hadronic function $H(x^2,0)$ is computed by averaging all data points of $H(x^2,P\cdot x)$ as follows
\be
\label{eq:H_relation}
H(x^2,0)=\frac{1}{\eta(P)}\int d\Omega_x\frac{x_\perp}{|x|}\frac{H(x^2,P\cdot x)}{\int_0^1du\,e^{i(u-\frac{1}{2})P\cdot x}\phi_\pi(x^2,u)}.
\ee
Here, we use a structure function $\phi_\pi(x^2,u)=1$ and treat $H^{\mathrm{LD}}(L,x^2,P\cdot x)$ from Eq.~(\ref{eq:H_LD}) similarly, following Eq.~(\ref{eq:H_relation}) 
to obtain $H^{\mathrm{LD}}(L,x^2,0)$. We fit $H^{\mathrm{LD}}(L,x^2,0)$ to lattice data at large $x$, using a fitting range of $|x|\in[1.5,3.0]$ fm for ensembles at near-physical pion masses 
and $|x|\in [1.0,2.5]$ fm for ensembles at $m_{\pi}\approx 340$ MeV to determine the parameters $m_V$ and $f_{\pi VV}$. 
These parameters, along with the fit quality, represented by $\chi^2/\text{d.o.f.}$, are listed in Table~\ref{tab:mrho_frhopipi} for all ensembles. The reasonable values of $\chi^2$/d.o.f. suggest 
that $H^{\mathrm{LD}}(L,x^2,P\cdot x)$ from Eq.~(\ref{eq:H_LD}) provides a good description of the lattice data. 
Due to the possible mixing between the vector meson and $\pi\pi$ state, the resulting $m_V$ is expected to be lower than the PDG value.
In Fig.~\ref{fig:VMD}, we illustrate the fit results for the 24D and 32D ensembles with different volumes. The plot shows that lattice results with a larger volume systematically shift upwards, 
as do the curves of $H^{\mathrm{LD}}(L,x^2,0)$, indicating the importance of including finite-volume corrections.

In the infinite volume, the corresponding hadronic function $H^{\mathrm{LD}}(x^2,P\cdot x)$ can be derived analytically. The Fourier transform leads to
\ba
\label{eq:LD_FFT}
\mathcal{F}_{\mu\nu}^{\mathrm{LD}}(Q,Q')&=& \int d^4x\, e^{-i\left(Q-\frac{P}{2}\right)\cdot x}\mathcal{H}_{\mu\nu}^{\mathrm{LD}}(x),
\nn\\
&=&i\,\varepsilon_{\mu\nu\alpha\beta}Q^\alpha P^\beta \mathcal{F}_{\pi^0\gamma^*\gamma^*}^{\mathrm{LD}}(-Q^2,-{Q'}^2),
\ea
where $Q'=P-Q$, and
\be
\mathcal{F}_{\pi^0\gamma^*\gamma^*}^{\mathrm{LD}}(-Q^2,-{Q'}^2)=\frac{f_{\pi VV}}{(Q^2+m_V^2)({Q'}^2+m_V^2)}.
\ee
The parameter $\tilde{\alpha} = F_\pi \mathcal{F}_{\pi^0\gamma\gamma}(0,0)$ can be approximated in the VMD model as $\tilde{\alpha}_{\mathrm{VMD}}=\frac{F_\pi f_{\pi VV}}{m_V^4}$. This ratio is expected to exhibit only mild dependence on the quark mass. This expectation is supported by the results in Table~\ref{tab:mrho_frhopipi}, where $\tilde{\alpha}_{\mathrm{VMD}}$ for the 48I and 64I ensembles differs from that of 24IH and 32IcH by only 8\%-12\%, despite the fact that $f_{\pi VV}$ varies by as much as 80\%.
Applying Feynman parameterization, we obtain
\ba
\mathcal{F}_{\pi^0\gamma^*\gamma^*}^{\mathrm{LD}}(-Q^2,-{Q'}^2)&=&\int_0^1 du\,\frac{f_{\pi VV}}{((1-u)Q^2+u{Q'}^2+m_V^2)^2}
\nn\\
&=&\int_0^1du\,\frac{f_{\pi VV}}{(\bar{K}^2+\bar{M}^2)^2}
\ea
with $\bar{K}=Q-uP$ and $\bar{M}^2=u(1-u)P^2+m_V^2$. 

Next, we express $\frac{f_{\pi VV}}{(\bar{K}^2+\bar{M}^2)^2}$ as
\be
\label{eq:FF_substitution}
\frac{f_{\pi VV}}{(\bar{K}^2+\bar{M}^2)^2}=\int d^4x\, G(x,\bar{M})e^{-i\bar{K}\cdot x},
\ee
with
\be
G(x,m)=\int \frac{d^4q}{(2\pi)^4}\,\frac{f_{\pi VV}\,e^{iq\cdot x}}{(q^2+m^2)^2}=\frac{f_{\pi VV}}{8\pi^2}K_0(m|x|).
\ee
Here, $K_n(z)$ represents the modified Bessel function of the second kind.
Substituting Eq.~(\ref{eq:FF_substitution}) into Eq.~(\ref{eq:LD_FFT}), we arrive at
\ba
&&\mathcal{F}_{\mu\nu}^{\mathrm{LD}}(Q,Q')
\nn\\
&=& i\int_0^1du\int d^4x\, G(x,\bar{M})\,\varepsilon_{\mu\nu\alpha\beta}\left(i\frac{\partial}{\partial x_\alpha}\right)  P_\beta e^{-i\bar{K}\cdot x}
\nn\\
&=&i \int_0^1du\int d^4x\,\varepsilon_{\mu\nu\alpha\beta}\left(-i\frac{\partial G(x,\bar{M})}{\partial x_\alpha}\right)  P_\beta e^{-i\bar{K}\cdot x}
\nn\\
&=&-\frac{f_{\pi VV}}{8\pi^2}\int_0^1du\int d^4x\,\varepsilon_{\mu\nu\alpha\beta}x_\alpha P_\beta \frac{\bar{M}}{|x|}K_1(\bar{M}|x|)  e^{-i\bar{K}\cdot x}.
\nn\\
\ea
On the other hand, using Eq.~(\ref{eq:F_munu}) and (\ref{eq:structure_func}), we derive 
\ba
\mathcal{F}_{\mu\nu}^{\mathrm{LD}}(Q,Q')&=& -\int d^4x\, \varepsilon_{\mu\nu\alpha\beta}x_{\alpha}P_{\beta} \int_{0}^{1}du\, e^{-i\bar{K}\cdot x} \times
\nn\\
&& \hspace{1cm}\phi_{\pi}^{\mathrm{LD}}(x^2,u)H^{\mathrm{LD}}(x^2,0).
\ea
By comparing these two equations, we obtain
\be
\frac{f_{\pi VV}}{8\pi^2}\frac{\bar{M}}{|x|}K_1(\bar{M}|x|) =\phi_{\pi}^{\mathrm{LD}}(x^2,u)H^{\mathrm{LD}}(x^2,0).
\ee 
This leads to the result
\be
H^{\mathrm{LD}}(x^2,0)=\frac{f_{\pi VV}}{8\pi^2}\int_0^1du\,\frac{\bar{M}}{|x|}K_1(\bar{M}|x|),
\ee
and
\be
\phi_\pi^{\mathrm{LD}}(x^2,u)=\frac{\bar{M}K_1(\bar{M}|x|)}{\int_0^1du\,\bar{M}K_1(\bar{M}|x|)}.
\ee
Given that $m_\pi^2\ll m_V^2$ and $\bar{M}^2\approx m_V^2$, $\phi_\pi^{\mathrm{LD}}(x^2,u)$ is nearly independent of $u$ and $|x|$. Therefore, we approximate $\phi_\pi^{\mathrm{LD}}(x^2,u)\approx 1$, which is 
consistent with $\phi_\pi^{\mathrm{VMD}}$ as defined in Eq.~(\ref{eq:model}).
The general hadronic function in infinite volume is given by
\be
H^{\mathrm{LD}}(x^2,P\cdot x)=\int_0^1 du\,e^{i(u-\frac{1}{2})P\cdot x}\phi_\pi^{\mathrm{LD}}(x^2,u)H^{\mathrm{LD}}(x^2,0).
\ee

By comparing the results obtained using either the finite-volume hadronic function $H^{\mathrm{LD}}(L,x^2,P\cdot x)$ or the infinite-volume function $H^{\mathrm{LD}}(x^2,P\cdot x)$, we can derive the TFF and
$a_{\mu}^{\pi^0\mathrm{-pole}}$, and subsequently determine the finite-volume effects.

\subsection{Contributions from disconnected diagrams}

For calculating the disconnected contribution from the light quark loop minus the strange quark loop, an efficient technique was developed by C. Michael et al. in 2007~\cite{Michael:2007vn}, utilizing the so-called one-end trick~\cite{Foster:1998vw,McNeile:2006bz}. This approach was later developed and extended as the split-even method in Ref.~\cite{Giusti:2019kff}. By combining this technique with low-mode averaging~\cite{DeGrand:2004qw} and all-mode averaging~\cite{Blum:2012uh}, A. Gérardin et al.~\cite{Gerardin:2023naa} demonstrated that the vector loop of the disconnected diagram can be efficiently computed.
It is worth noting that a different method was used in an earlier work by RBC-UKQCD Collaboration~\cite{Blum:2015you} to calculate the disconnected contribution in hadronic vacuum polarization. In that approach, the vector loop was computed using sparse random sources combined with full-volume low-mode averaging, which also proved to be highly effective.
In this work, we employ the split-even method to make use of existing data efficiently. 
The essential feature of both the one-end trick and the split-even method is that they allow the light-minus-strange loop contribution to be expressed as
\ba
L(x)&=&\operatorname{Tr}[\Gamma D^{-1}_l(x,x)-\Gamma D^{-1}_s(x,x)]
\nn\\
&=&(m_s-m_l)\sum_{y\in L^3}\operatorname{Tr}[\Gamma D^{-1}_l(x,y)D^{-1}_s(y,x)]
\nn\\
&=&(m_s-m_l)\sum_{y\in L^3}\operatorname{Tr}[\Gamma D^{-1}_l(x,y)\gamma_5 D^{-1\dagger}_s(x,y)\gamma_5]
\nn\\
\ea
where $\Gamma$ represents the gamma matrix, $D^{-1}_{l/s}(x,y)$ is the light/strange quark propagator, and $m_{l/s}$ is the bare light/strange quark mass. 
Using $U(1)$ stochastic sources $\xi^{(r)}(y)$ with $r=1,2,\cdots, N_r$, the propagators $\phi_{l/s}^{(r)}(x)$ can be written as
\be
\phi_{l/s}^{(r)}(x)=\sum_{y\in L^3} D^{-1}_{l/s}(x,y)\xi^{(r)}(y).
\ee 
The one-end trick allows us to construct the loop contribution as
\be
L(x)=\frac{m_s-m_l}{N_r}\sum_{r=1}^{N_r}\operatorname{Tr}[ \Gamma\phi_l^{(r)}(x) \gamma_5 \phi_s^{(r)\dagger} (x)\gamma_5].
\ee

\begin{figure}[htb]
\centering
\includegraphics[width=0.48\textwidth,angle=0]{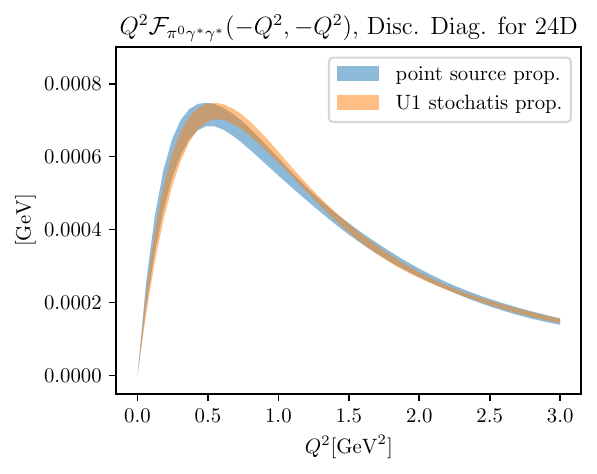}
\caption{Disconnected contributions to $Q^2\mathcal{F}_{\pi^0\gamma^*\gamma^*}(-Q^2,-Q^2)$. 
The results are computed using one-end trick with both point-source propagators and $U(1)$ stochastic propagators for comparison.
Results are presented for the 24D ensemble with the VMD parameterization $\phi_\pi(x^2,u)=1$ as an example.
Different parameterizations of $\phi_\pi(x^2, u)$ yield highly consistent results.
}
\label{fig:Fqq_disc}
\end{figure}

For the 24D ensemble, we used $N_r=64$ stochastic-source propagators to compute the loop function $L(x)$. Additionally, we have already computed 1024 point-source propagators per configuration to construct the connected diagrams. Since these propagators are available, we used them for free to build $L(x)$. A comparison is shown in Fig.~\ref{fig:Fqq_disc} for the computation of TFF, revealing comparable uncertainties. For $a_{\mu}^{\pi^0\mathrm{-pole}}$, the results for 24D are
\be
a_{\mu}^{\mathrm{disc}}\times10^{11}\Big|_{\mathrm{24D}}=\begin{cases}
1.70(23), & U(1)\mbox{-stochastic}\\
1.92(27), & \mbox{point-source}
\end{cases}.
\ee
Given that the number of stochastic-source propagators is much smaller than that of the point-source propagators, the combination of stochastic-source propagators and the one-end trick offers a more efficient approach. However, in our calculation, since we already accumulated thousands of point-source propagators per configuration, we utilized them for the calculation of the disconnected diagrams.

For the 64I ensemble, we computed 2048 point-source propagators per configuration, allowing for a more accurate determination of the disconnected diagram. We find
\be
a_{\mu}^{\mathrm{disc}}\times10^{11}\Big|_{\mathrm{64I}}=1.59(10),
\ee
which is in good agreement with the 24D results. This suggests that, for the domain wall fermion ensembles used in this calculation, lattice artifacts in the disconnected diagram 
are not significant relative to statistical uncertainties.

\begin{figure}[htb]
\centering
\includegraphics[width=0.48\textwidth,angle=0]{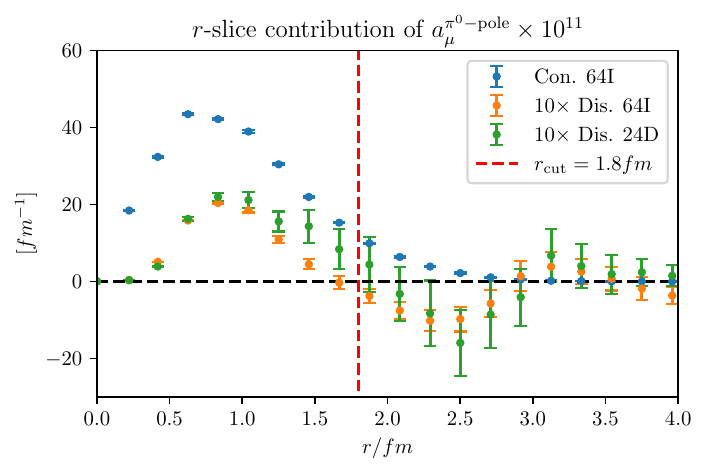}
\caption{Comparison of the 64I disconnected contribution (orange) with the 24D disconnected (green) and 64I connected contributions (blue).
}
\label{fig:amu_disc_diag}
\end{figure}

In Fig.~\ref{fig:amu_disc_diag}, we compare the 64I disconnected contribution with the 24D disconnected and 64I connected contributions.
By "disconnected contribution", we refer to the case where one form factor (labeled $A$) uses the connected contribution while the other (labeled $B$) uses the disconnected contribution. For the "connected contribution", both form factors $A$ and $B$ use the connected contribution. We introduce a parameter $r$ in the spacetime integral for form factor $B$, while the form factor $A$ is fully integrated over the entire spacetime.
The figure shows the integrand as a function of $r$, and 
integrating over $r$ provides the contribution to $a_{\mu}^{\pi^0\mathrm{-pole}}$. From the figure, we confirm that the 64I and 24D disconnected contributions are consistent with each other and are both more than 10 times smaller than the 64I connected contribution. At large $r$, the results vanish but become very noisy, so we introduce a truncation for the disconnected contribution at $r_{\mathrm{cut}}=1.8$ fm. The residual truncation effects are neglected, as the disconnected contributions are significantly smaller than the connected ones.

It is worth noting that the disconnected contributions computed here have an opposite sign relative to previous lattice calculations~\cite{Feng:2012ck,Gerardin:2016cqj,Gerardin:2019vio,Gerardin:2023naa,ExtendedTwistedMass:2023hin,Koponen:2023zle,Christ:2022rho}. 
Although these contributions are relatively small, they become statistically significant when combined with the connected part, 
making it important to determine the correct sign for the disconnected diagrams.

\subsection{Unphysical pion mass effects}

\begin{figure}[htb]
\centering
\includegraphics[width=0.48\textwidth,angle=0]{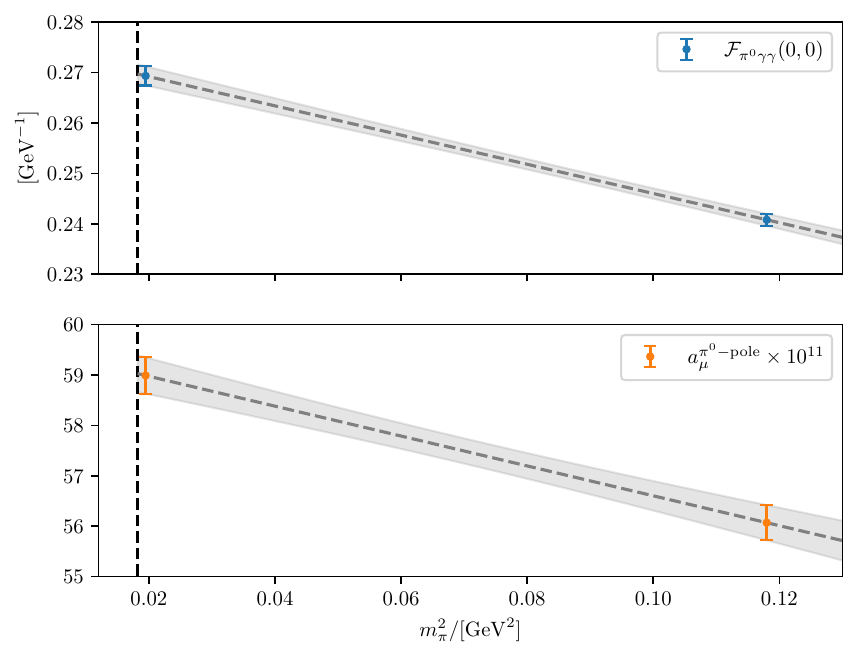}
\caption{Pion mass dependence of $\mathcal{F}_{\pi^0\gamma\gamma}(0,0)$ (upper panel) and the connected contribution to $a_{\mu}^{\pi^0\mathrm{-pole}}$ (lower panel),
shown for the two ensembles 48I and 32IcH.
}
\label{fig:mass_dep}
\end{figure}

\begin{figure}[htb]
\centering
\includegraphics[width=0.48\textwidth,angle=0]{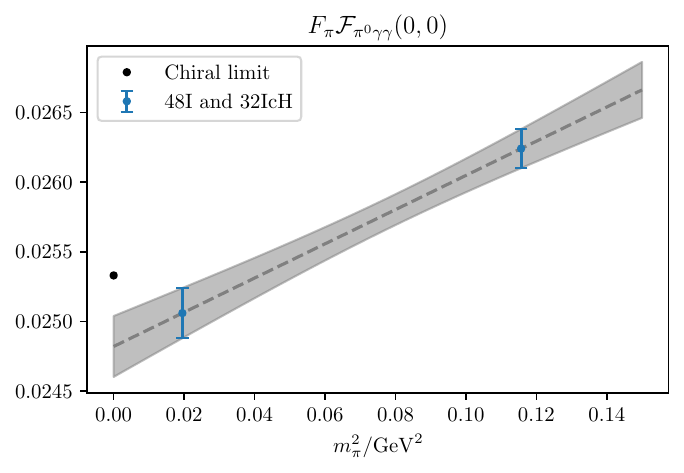}
\caption{Pion mass dependence of $F_\pi\mathcal{F}_{\pi^0\gamma\gamma}(0,0)$,
shown for the two ensembles 48I and 32IcH.}
\label{fig:mass_dep_1}
\end{figure}

\begin{table}[htbp!]
    \centering
    \begin{tabular}{c| c c c c }
        \hline
        Ens &  $m_\pi$ [MeV] & $\mathcal{F}_{\pi^0\gamma\gamma}$ [GeV$^{-1}$] & $F_\pi$ [GeV] &  $F_\pi \mathcal{F}_{\pi^0\gamma\gamma}$ \\
        \hline
        48I & $139.7(3)$ & 0.2697(19) &  0.0929(1) & 0.02506(18) \\
        32IcH & $340.1(1.2)$ & 0.2414(12) & 0.1087(2) & 0.02624(14) \\
        \hline
    \end{tabular}
    \caption{Lattice results used to compile Figs.~\ref{fig:mass_dep} and \ref{fig:mass_dep_1}. $\mathcal{F}_{\pi^0\gamma\gamma}$ stands for $\mathcal{F}_{\pi^0\gamma\gamma}(0,0)$.}
    \label{tab:mass_extrapolation}
\end{table}

In Fig.~\ref{fig:mass_dep} we present the pion mass dependence of $\mathcal{F}_{\pi^0\gamma\gamma}(0,0)$ (upper panel) and the connected contribution to $a_{\mu}^{\pi^0\mathrm{-pole}}$ (lower panel). 
The two ensembles used here, 48I and 32IcH, have similar lattice spacings, and finite-volume corrections have been applied to these results. 
In Fig.~\ref{fig:mass_dep_1} we also present the pion dependence of the normalized TFF $F_\pi\mathcal{F}_{\pi^0\gamma\gamma}(0,0)$, again using only connected contributions. 
The corresponding results are summarized in Table~\ref{tab:mass_extrapolation}.
Our current dataset includes one near-physical point and another at $m_\pi\approx 340$ MeV.
We perform a simple linear fit to the two available data points and obtain the mass corrections to both $\mathcal{F}_{\pi^0\gamma\gamma}(0,0)$ and $a_\mu^{\pi^0\text{-pole}}$. These corrections, listed in Table~\ref{tab:results_F00} and Table~\ref{tab:syst_budget}, are small compared to other systematic uncertainties, as expected, since the relevant ensembles are already very close to the physical pion mass.

Apart from the mass correction, the slope of $F_\pi \mathcal{F}_{\pi^0\gamma\gamma}(0,0)$ can determine the chiral low-energy constant (LEC) $C_7^{Wr}$ in the chiral perturbation theory~\cite{Bijnens:1988kx,Goity:2002nn,Kampf:2009tk,Gerardin:2019vio} via
\ba
F_\pi \mathcal{F}_{\pi^0\gamma\gamma}(0,0) = \tilde\alpha - \frac{64}{3} C_7^{Wr} m_\pi^2.
\ea
From the slope, we extract the connected contribution to the LEC, $C_7^{Wr,\mathrm{conn}} = -0.61(11) \times 10^{-3}\mbox{ GeV}^{-2}$. As shown in Table~\ref{tab:mass_extrapolation}, much of the pion-mass dependence in $F_\pi$ and $\mathcal{F}_{\pi^0\gamma\gamma}(0,0)$ cancels out in the product, rendering $C_7^{Wr}$ particularly sensitive to subleading effects.
Noting that the disconnected contribution to $F_\pi \mathcal{F}_{\pi^0\gamma\gamma}(0,0)$ vanishes in the chiral limit, 
we therefore can calculate the slope using the data from the 64I ensemble alone, obtaining $C_7^{Wr,\mathrm{disc}} = -0.77(7) \times 10^{-3}\mbox{ GeV}^{-2}$. 
Combining both contributions, we arrive at the total result $C_7^{Wr} = -1.38(13) \times 10^{-3}\mbox{ GeV}^{-2}$.
Our result differs from the value reported in Ref.~\cite{Gerardin:2019vio}, $C_7^{Wr} = 0.16(18) \times 10^{-3}\mbox{ GeV}^{-2}$. This discrepancy could be partly due to the sizable disconnected contribution entering with the opposite sign in the earlier determination.
Note that the extraction of $C_7^{Wr}$ in this work should be interpreted with caution,
as it is subject to following potential systematic uncertainties that are difficult to quantify.
The quoted uncertainty for $C_7^{Wr,\mathrm{conn}}$ above is purely statistical and does not account for the following sources of systematics:
(1) The data point at $m_\pi \approx 340~\mathrm{MeV}$ may lie outside the regime where 
a linear dependence of $F_\pi \mathcal{F}_{\pi^0\gamma\gamma}(0,0)$ on $m\pi^2$ is valid;
(2) The finite-volume correction estimated using the VMD model is much larger for the 32IcH ensemble (3.2\%, see Table~\ref{tab:results_F00}) 
than for the 48I ensemble (0.4\%); (3) The time separation between the pion interpolating operator and the two vector currents is shorter for 32IcH 
($\Delta t = 0.66~\mathrm{fm}$) compared to 48I ($\Delta t = 1.37~\mathrm{fm}$), which may enhance excited-state contamination.
These issues can be addressed in future studies by including more accurate data at a range of pion masses or by directly computing the mass derivative of the TFF.
Finally, we stress that these systematic uncertainties affect only the determination of $C_7^{Wr}$,
but do not impact the accuracy of our main results. The opposite sign of the disconnected contribution, as observed in this work,
is relevant for the determination of $C_7^{Wr}$ itself and could potentially impact the chiral prediction for the $\pi^0\to\gamma\gamma$ decay width.

\begin{table}[htbp!]
    \centering
    \begin{tabular}{c| c c c c | c c}
        \hline
    & \multicolumn{5}{c}{$a_{\mu}^{\pi^0\mathrm{-pole}}\times 10^{11}$} \\
        \hline
        Ens &  conn. diag. & FV corr. & mass corr. & disc. diag. & total\\
        \hline
        24D & $63.23(39)$ & 0.77(6) & 0.06(1) & 1.92(27) & 65.65(40) \\
        32D & $64.34(47)$ & 0.057(6) & 0.06(1) & - & 66.05(48) \\
        32Df & $58.88(63)$ & 0.80(11) &  0.06(1) & - & 61.33(48) \\
        48I & $58.83(36)$ & 0.16(1) &  0.034(6) & - & 60.61(37)\\
        64I & $59.08(47)$ & 0.19(2)  &  0.028(6) & 1.59(10) & 60.88(48)\\
        24DH & $57.18(63)$ & 0.26(1) & - & - & - \\
        24IH & $47.21(30)$ & 7.67(24) & - & - & - \\
        32IcH & $53.94(34)$ & 2.13(8) & - & - & - \\
        \hline
    \end{tabular}
    \caption{Summary of connected contribution to $a_{\mu}^{\pi^0\mathrm{-pole}}$, along with finite-volume corrections, unphysical pion mass corrections, and disconnected contributions across different ensembles.}
    \label{tab:syst_budget}
\end{table}

\subsection{Excited-state contamination}

\begin{figure}[htbp!]
    \centering
    \includegraphics[width=0.48\textwidth]{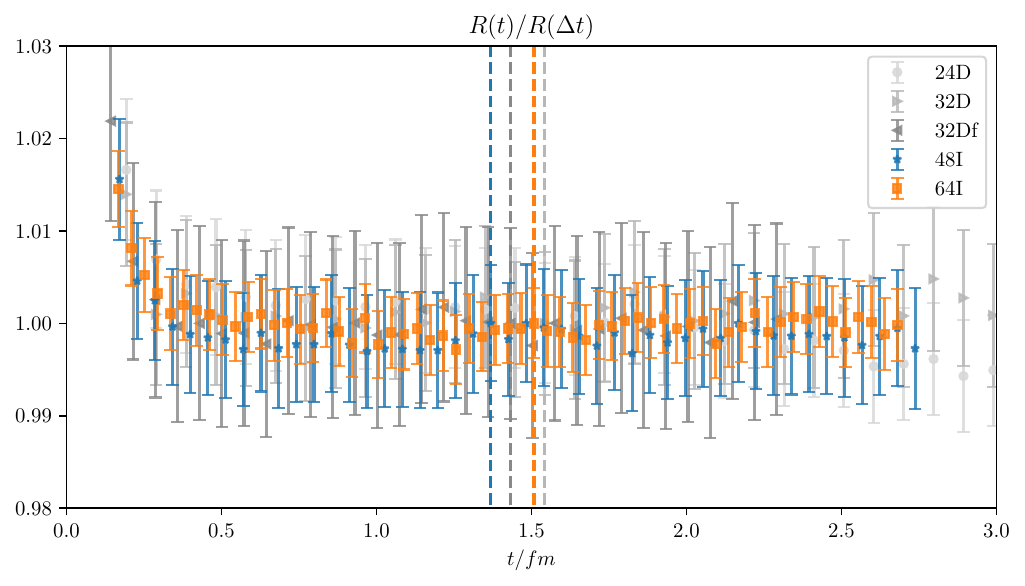}
\includegraphics[width=0.48\textwidth]{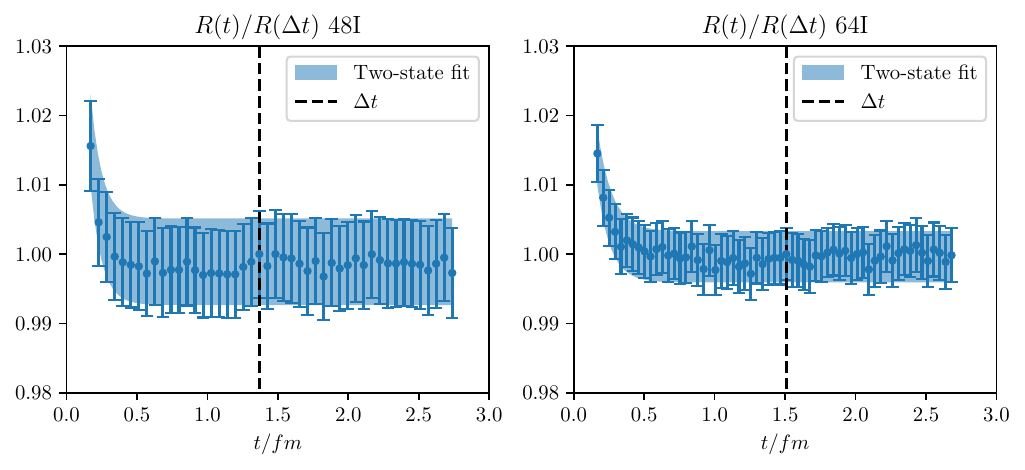}
    \caption{Upper panel: $R(t)/R(\Delta t)$ as a function of $t$ for all the ensembles with near-physical pion masses. lower panel: Two-state fit for $R(t)/R(\Delta t)$, with 48I and 64I used as examples.}
     \label{fig:R_t}
\end{figure}

To estimate the contamination from excited states, we define the ratio
\be
R(t)=\frac{\langle O_\pi(2t)\bar{O}_\pi(0)\rangle}{\cosh(m_\pi(2t-T/2))},
\ee
where the wall-source pion interpolating operator $\bar{O}_\pi$ is used to create a pion from the vacuum at the source, while
$O_\pi$ is used to annihilate the pion at the sink.
In the absence of excited-state contamination, this ratio should be constant. 
In the upper panel of Fig.~\ref{fig:R_t}
we plot $R(t)/R(\Delta t)$ as a function of $t$. Since the wall-source operator 
$\bar{O}_\pi$ and $O_\pi$ have a good overlap with the $\pi$ ground state, we achieve ground-state saturation 
for $t \sim 0.5$ fm. As the pion correlator with zero momentum 
is not affected by the signal-to-noise problem,
we conservatively choose $\Delta t$ to be around 1.5 fm, 
as indicated by the vertical dashed lines in Fig.\ref{fig:R_t}.

To estimate the magnitude of excited-state contamination, we perform a two-state fit and 
present the results for the 48I and 64I ensembles as examples in the lower panel 
of Fig.~\ref{fig:R_t}. 
From the two-state fit, we determine
\be
\epsilon=\left(\frac{R(\Delta t)}{R(\infty)}-1\right)^{\frac{1}{2}}=\begin{cases}
0.7(0.5)\times 10^{-4}, & \mbox{for 48I}\\
1.3(1.7)\times10^{-4}, & \mbox{for 64I}
\end{cases}.
\ee
It is important to note that when determining $\epsilon$, both $R(\Delta t)$ and $R(\infty)$ are obtained from the fit. 
However, when compiling the ratio $R(t)/R(\Delta t)$ and the fitting curve in Fig.~\ref{fig:R_t}, the denominator $R(\Delta t)$ 
uses only the central value of the raw data, without error propagation.
The final values of $\epsilon$ are on the order of $10^{-4}$, indicating that at $\Delta t$, the excited-state contamination is negligible.

\subsection{Summary of systematic effects}

\begin{figure}[htbp!]
    \centering
    \includegraphics[width=0.46\textwidth]{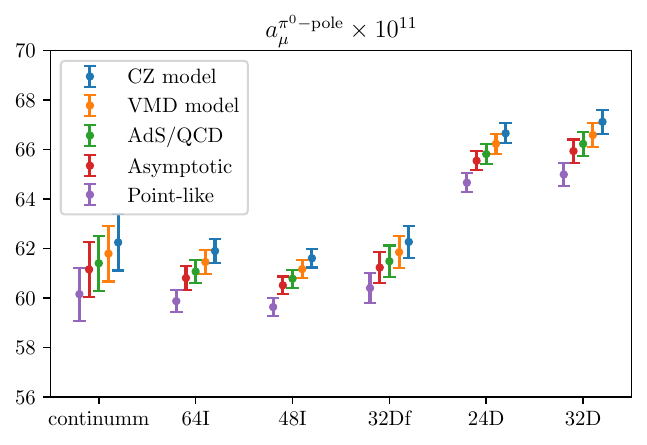}
    \caption{Lattice results of $a_{\mu}^{\pi^0\mathrm{-pole}}$ as a function of $a^2$ using different parameterizations of $\phi_\pi(x^2,u)$. After a continuum extrapolation with a linear form of $a^2$, the uncertainty increases significantly, leaving the variance from the parameterizations secondary important.}
     \label{fig:amu_diff_phi}
\end{figure}

Table~\ref{tab:syst_budget} provides detailed information on the dominant connected contribution to $a_{\mu}^{\pi^0\mathrm{-pole}}$, along with finite-volume corrections, unphysical pion mass corrections, 
and disconnected contributions.
The disconnected contributions from ensembles 24D and 64I 
show good agreement, and we apply the 64I result as the disconnected correction for the other ensembles.
Fig.~\ref{fig:amu_diff_phi} illustrates that the five different parameterizations of $\phi_\pi(x^2,u)$ yield varying results for $a_{\mu}^{\pi^0\mathrm{-pole}}$. Following a continuum extrapolation using a linear form 
in $a^2$, the uncertainty increases substantially, making the differences among parameterizations less significant.
The values listed in Table~\ref{tab:syst_budget} represent the average of the maximum and minimum values across the five parameterizations.

\end{document}